\documentclass[aps,prd,amsmath,showpacs,showkeys,preprint]{revtex4}
\usepackage{amsmath}
\usepackage[cp1251,koi8-r]{inputenc}
\usepackage[english]{babel}
\usepackage{graphics}
\usepackage[dvips]{color}
\usepackage{epsfig}
\newcommand{\be}{\begin{equation}}
\newcommand{\ee}{\end{equation}}
\newcommand{\bn}{\begin{eqnarray}}
\newcommand{\en}{\end{eqnarray}}
\newcommand{\bd}{\begin{displaymath}}
\newcommand{\ed}{\end{displaymath}}
\newcommand{\bnn}{\begin{eqnarray*}}
\newcommand{\enn}{\end{eqnarray*}}
\newcommand{\bs}{\begin{subequations}}
\newcommand{\es}{\end{subequations}}
\newcommand{\ba}{\begin{aligned}}
\newcommand{\ea}{\end{aligned}}
\def\Ref#1{(\ref{#1})}
\def\Journal#1#2#3#4#5#6{#1 \ #2 \ #3 \  #4 \ #5 \ #6}

\begin{document}
\inputencoding{cp1251}
\title{Relaxational Singularities of Human Motor System at Aging
\\ Due to Short-Range and Long-Range Time Correlations}

\author{\firstname{Renat~M.}~\surname{Yulmetyev}$^{1,2}$}
\email{rmy@theory.kazan-spu.ru}

\author{\firstname{David~E.}~\surname{Valliancourt}$^{3}$}

\author{\firstname{Fail~M.}~\surname{Gafarov}$^{2}$}

\author{\firstname{Sergey~A.}~\surname{Demin}$^{1,2}$}

\author{\firstname{Oleg~Yu.}~\surname{Panischev}$^{1,2}$}

\author{Peter H\"anggi$^4$}

\affiliation{$^1$Department of Physics, Kazan State University,
420008 Kazan, Kremlevskaya Street, 18 Russia \\ $^2$Department of
Physics, Kazan State Pedagogical University, 420021 Kazan,
Tatarstan Street, 2 Russia \\ $^3$School of Kinesiology,
University Illinois at Chicago, Chicago, Illinois, 60608 USA \\
$^4$Institute of Physics, University of Augsburg,
Universit\"atsstra\ss e 1, D-86135 Augsburg, Germany}

\begin{abstract}
In this paper we study the relaxation singularities  of human
motor system at aging. Our purpose is to examine the structure of
force output variability as a function of human aging in the time
and frequency domains. For analysis of experimental data we have
developed  here the statistical theory of relaxation of force
output fluctuation with taking into account the effects of two
relaxation channels. The first of them contains the contribution
of short-range correlation whereas other relaxation component
reflects the effect of long-range correlation. The analysis of
experimental data shows, that the general behavior of relaxation
processes  at human aging is determined by a complicated
combination and nonlinear interactions  two above stated
relaxation processes as a whole.
\end{abstract}
\pacs{05.45.Tp, 45.05.+x, 87.19.St, 89.75.-k}

\keywords{Human aging, discrete non-Markov processes, statistical
memory, short-time and long-time relaxation channels}

\maketitle

\section{Introduction}
Recent studies the aging of human neuromuscular system has focused
on application of central theory regarding the age relating with
the structure of behavioral and physiological variability due to
loss of "complexity" \cite{Lip,Vall1,Vall2}. The term  of
"complexity" is connected  to the broader concept including the
fractals and dynamics in disease and aging. The concept consider
behavioral and physiological  changes of the system's due to
aberrations in their time and frequency structure. However it is
necessary to specify that less attention has been given to the
neuromuscular changes associated with the time-dependent and
dynamic peculiarities of force control till now.

In this study, we examine the time and frequency structure of
healthy adult humans to determine a relaxation singularities,
arising with aging of human motor systems. It is well known that
aging tends to induce a range of performance decrements in human
motor system. A common finding is that the amount of motor
variability increases in the healthy aging adult over a broad set
of tasks \cite{Gol,Zai,tra,Vall3}.

Although consideration of the amount of variability is an
important indicator of the aging neuromuscular system, the
structure of motor variability also provides significant insight
into system organization and human motor control
\cite{Gold,Hau,Kap,New}. The presence of short-time and long-range
correlations in neuromuscular fluctuations in healthy people has
implications for understanding and mathematical modelling of
neuroautonomic regulation. Here we have applied correlation
analysis to assess the effects of physiological aging on
correlation behavior of human motor system. The purpose of this
paper was to consider the effects of aging and task demand on the
relaxation structure of the signals of force output variability
from the point of view of modern statistical physics. We consider
the force-time series data as a discrete stochastic process and
apply the  statistical theory and the information measure of
memory of non-Markov random processes in complex systems
\cite{main,eqpap,ycard,jetp}. Therefore we can use the notions,
based on the theory of chaos and information, in our analysis of
complex systems. This allows  to apply in our consideration
representations and notions on statistical short-time and
long-time memory, information measures of memory, relaxation and
kinetic parameters from statistical theory
\cite{Shur1,Shur2,Yulm3,Shur3,Shur4,PhysAInt,PhysaSens,PhysaGait}.

\section{Methods}

\subsection{ Participants }
A total of 29 participants were assigned to three different age
groups: young group (N=10; range: 20 - 24 yrs; mean: 22 + 1 yrs; 5
females and 5 males), old group (N=9; range: 64 - 69 yrs; mean: 67
+ 2 yrs; 4 females and 5 males), and older-old group (N=10; range:
75 - 90 yrs; mean: 82 + 5 yrs; 5 females and 5 males).  All of the
participants were right hand dominant.  The participants were
familiarized  with the purpose of the experiment and all
participants gave informed consent to all experimental procedures,
which were approved by the local Institutional Review Board.

The three age groups consisted of moderately active individuals.
Persons who were highly active were excluded from the study. Also,
elderly persons who were considered frail were excluded from the
study. Twelve of the participants in the two elderly groups were
taking medication for the treatment of high blood pressure. The
distribution of persons taking medication for high blood pressure
was 8 in the older-old group and 4 in the old group. None of the
elderly participants reported having a neuromuscular or
neuropsychiatric disease, nor did any of the participants have
diabetes. Also, twelve of the elderly participants reported having
arthritis and each were taking medication for the condition.  The
distribution of persons reporting arthritis of the hand was 7 in
the older-old group and 5 in the old group.  All participants
remained on their normal medications during testing.

\subsection{ Apparatus }

Participants were seated in a chair with their dominant forearm
resting on a table (75 cm in height).  The participant's dominant
hand was pronated and lay flat on the table with the digits of the
hand comfortably extended.  The setup constrained the wrist and
the third, fourth and fifth phalanges from moving.  The elbow
position remained constant throughout the experimental session.
Through abduction, the participant's lateral side of the index
finger contacted the load cell (Entran ELFS-B3, New Jersey), 1.27
cm in diameter, which was fixated to the table.  The load cell was
located 36 cm from the participant's body midline.  Analog output
from the load cell was amplified through a Coulbourn Type A
(Strain gauge Bridge) S72-25 amplifier at an excitation voltage of
10 V and a gain of 100.  A computer controlled 16-bit A/D
converter sampled the force output at 100 $Hz$.  The smallest
increment of change in force the A/D board could detect was .0016
N.  The force output was displayed on a video monitor located 48.6
cm from the participants' eyes and 100 cm from the floor.
According to previous work from laboratory (D.A.V.), the
display-to-control gain was set at 20 pixels per 1 N change in
force for each participant.

\subsection{ Procedures }

During the initial portion of the experiment, the participant's
maximum voluntary contraction (MVC) was estimated consistent with
previous work \cite{Vall3}. Participants abducted their index
finger against the load cell with maximal force for three
consecutive 6 $s$ trials.  A 60 $s$ rest period was provided for
each participant between each MVC trial. In each MVC trial, the
mean of the greatest ten force samples was calculated.  The means
obtained from three trials were averaged to provide an estimate of
each participant's MVC. Participants adjusted their level of force
output to match a red target line (1 pixel thick) on the video
monitor.  Participants viewed online feedback of their performance
in the form of a yellow force-time trajectory that moved from left
to right in time across the video monitor.  They were instructed
to match the yellow trajectory line to the red horizontal target
line throughout each trial, and to minimize all deviations of the
yellow line from the red line.

Participants produced force at a constant force target.  The
constant target was a horizontal line displayed across the center
of the video monitor.   Participants produced force at 5, 10, 20
and 40 \% of their MVC under the constant force condition for two
consecutive 25 $s$ trials at each force level. A rest period of
100 $s$ was provided between each force trial.  The order of the
force and target conditions was randomized across all
participants.

\section{Data Analysis}
The force-time series data were conditioned by the following
methods. First, force data were digitally filtered by using a
first-order Butterworth filter with a low-pass cutoff frequency of
20 $Hz$. All data processing and subsequent time and frequency
analysis were performed by using software written in Matlab. The
analysis of force output concentrated on two primary problems: 1)
calculation of the information measure of memory of force
variability; 2) study of the correlation and relaxation structure
of force time variability.

\subsection{The information measure of memory for force variability.}

For comparison the relaxation time scales of the initial TCF
$a(t)$ and memory functions of the $i$th order $M_i(t)$ (they will
be described below in Eqn. (3)) we use here the first measure of
memory that is the statistical non-Markovian parameter. Originally
the non-Markovian parameter, characterizing the degree of
non-Markovity of an arbitrary relaxation process, was introduced
for analyzing the irreversible phenomena in condensed matters
\cite{Shur1,Shur2}. The relaxation times of initial TCF (the
existence duration of correlations in considered system) and
memory functions of the $i$th order (the duration of existence of
memory) can be determined as follows: $\tau_a=\Delta t
\sum_{j=o}^{N-1}a(t_j),\tau_{M_1}=\Delta t
\sum_{j=o}^{N-1}M_1(t_j)$. The simplest criterion for the
quantitative estimation of non-Markovity in the given relaxation
process is determined as: $\varepsilon_1=\tau_a/\tau_{M_1}$. When
$\varepsilon_1\gg1$, the relaxation time of the memory function of
the first order is much smaller than the relaxation time of the
initial TCF. In this case the process is characterized with a very
short memory, in the limit $\varepsilon \rightarrow \infty$ it is
a Markovian process. Particularly, our work \cite{Yulm3} proposes
the way of the transformation from non-Markovian kinetic equations
to Markovian in case, when the non-Markovian parameter tends to
infinity. Decreasing the parameter $\varepsilon_1$ determines the
relative memory lengthening and strengthening of non-Markovian
time effects. Thus, the presented quantitative criterion
characterizes the degree of non-Markovity and strength of memory
in the underlying relaxation process.

Later \cite{Shur3,Shur4} a conception of non-Markovian parameter
spectrum ${\varepsilon}$ and markovization depth for
nonequilibrium processes in disordered condensed matter was
introduced. These parameters are related to fundamental properties
of the system as well as the memory function, the memory life time
and demarkovization of the process by means of the initial TCF.
The spectrum of non-Markovian parameter
$\{\varepsilon\}=\{\varepsilon_1,\varepsilon_2,\ldots,\varepsilon_{n-1}\}$
is a set of dimensionless $\varepsilon_i$ values:

\be
\varepsilon_1=\tau_a/\tau_{M_1},~\varepsilon_2=\tau_{M_1}/\tau_{M_2},~\ldots,~\varepsilon_{n-1}=
\tau_{M_{n-1}}/\tau_{M_n}.\nonumber \ee Here $\tau_i$ is a
relaxation time of memory function of $i$th order, the number $i$
defines the relaxation level.

The quantitative information measure of memory of force
variability was assessed by calculating the statistical spectrum
of non-Narkovity parameter (NMP). The equation: \be
\varepsilon_i(\omega)=\left \{ \frac{\mu_i(\omega)}{\mu_i(\omega)}
\right \}^\frac{1}{2}, \label{a} \ee
 where $i=1,2,3, ...$ is the number
of relaxation levels in kinetic description \cite{main} of the
time series, allow to calculate NMP $\varepsilon_i$ from the power
spectra of memory function of $i$th order $\mu_i(\omega)$. The
mathematical procedure for a finding of $\mu_i(\omega)$ consist in
the calculation: \be \mu_i(\omega)=\tau^{2}\left|\sum_{j=0}^{n-1}
M_i(j \tau) \cos(j \omega \tau)\right|^2, \label{b}\ee where
$M_i(t)$ $(t=j\tau)$ is the memory function of $i$th order,
$\omega=2\pi/\tau$, $\tau=0,01 s$ is the discretization time, $n$
is the number of signals in time series. Memory functions, phase
portraits of the dynamical orthogonal variables, and set of
dynamical orthogonal variables of junior orders were calculated by
the methods of well-known statistical theory of stochastic
discrete non-Markov processes in complex systems with applications
to cardiology, seismology, physiology etc. The full details of the
theory have been described elsewhere \cite{main,eqpap,ycard,jetp}.

\subsection{The correlation and relaxation analysis of force variability.}
The structure of correlation, relaxation and memory processes in
force variability will be examined by using  time and frequency
analysis of correlation and studied relaxation processes.
Correlation and relaxation rates for the
 short-range $R_i^ {(s)}$
($R_i^ {(s)}=|\lambda_i|, i=1,2$) and long-range $R_i^
{(l)}$($R_i^ {(l)}$=$|\Lambda_i|^{1/2}, i=1,2$) relaxation
(correlation)  were used to consider the force signal. We shall
calculate a general relaxation time in a frame of relaxation
theory developed below.

Let's  consider the  structure of the initial time correlation
function $a(t)$ of the force output time series signals. It is
convenient to consider the chain of interconnected equations for
the TCF $a(t)$ in a frame of well-known Zwanzig'-Mori's kinetic
theory \cite{main,eqpap,ycard,jetp}:

\be \left\{
\begin{aligned}
\frac {da (t)} {dt} &=\lambda_1 a
(t) - \Lambda_1 \int_{0}^{t} a(t- \tau)M_1(\tau)d\tau,\\
\frac {dM_1 (t)} {dt} &=\lambda_2 M_1 (t) - \Lambda_2 \int_{0}^{t}
M_1(t- \tau)M_2(\tau) d \tau,\\
\ldots \ldots &\ldots \ldots \ldots \ldots \ldots \ldots \\
\end{aligned}
\right.\label{f1}\ee Here, the parameter $|\lambda_i|$ and
$|\Lambda_i|^{1/2}$ define the local and memory channel relaxation
(correlation) rates on the $i$th relaxation levels,
correspondingly. Parameters $\lambda_i$ has the dimension of
relaxation rate, and $\Lambda_i$ has a dimension of a square of
frequency.

The structure of correlation and relaxation processes in force
variability will be examined by using  time and frequency analysis
of correlations and studied relaxation processes. Correlation and
relaxation rates for the
 short-range $R_i^ {(s)}$
($R_i^ {(s)}=|\lambda_i|, i=1,2$) and long-range $R_i^
{(l)}$($R_i^ {(l)}$=$|\Lambda_i|^{1/2}, i=1,2$) relaxation
(correlation) were used to consider the force signal. We shall
calculate a general relaxation time in a frame of relaxation
theory developed below.

Let's consider a partial solution of system (3) when we can use
the time-scale invariance idea in nonequilibrium statisticalw
physics of condensed matter \cite{TPre,JetpL,JPhys,Ball}: \be
M_2(t)=M_1(t) \label{f2}\ee on the second relaxation level. It
means the approximate equality of memory life-times on the first
and second relaxation levels. The similar condition  is carried
out in many concrete cases. The condition (4) apply on our
situation as we shall see later. Using the Laplace transform on
the normalized TCF $a(t)$:
\bn \tilde a(s)=\int_0^\infty dt e^{-st} a(t), \ \
M_i(s)=\int_0^\infty dt e^{-st} M_i(t), \en
 memory functions $M_1(t)$ and
$M_2(t)$, one can obtain solution of Eqns. \Ref{f1}:
\bn
\tilde M_1(s)=\frac{1}{2\Lambda_2} \{ -(s+R_2)\pm \sqrt{(s+R_2)^2+4\Lambda_2} \}, \\
\nonumber \tilde a(s)= \{  s+R_1+\frac{\Lambda_1}{2 \Lambda_2} [
-(s+R_2)+ \\ \nonumber \sqrt{(s+R_2)^2+4\Lambda_2} ]\}^{-1} , \en
where $R_{i}=R_{i}^{(s)}= {|\lambda_i|}$, i=1, 2. If we determine
the general relaxation time $\tau_R$ by the relation $\tau_R=\Re
\lim_{s \to 0} \tilde a(s)$, where $\Re$ means real part, the
resulting general equations for $\tau_R$ consequently can be
written in the following way:
\be \tau_R=\{ R_1+\frac{\Lambda_1}{2 \Lambda_2}[\sqrt{4
\Lambda_2+R_2^2}-R_2]\}^{-1}, \label{f7} \ee
if $4 \Lambda_2+R_2^2 \geq 0$. Then we receive the formula:
\be \tau_R=\frac{R_1-\frac{\Lambda_1}{2 \Lambda_2}R_2}
{(R_1-\frac{\Lambda_1}{2 \Lambda_2}R_2)^2+\frac{\Lambda_1^2}{4
\Lambda_2^2}|4\Lambda_2+R_2^2|}, \label{f8}\ee
for $\Lambda_2<0$, $4\Lambda_2+R_2^2<0$.

The equations \Ref{f7}, \Ref{f8} are our new theoretical results
and they allow to calculate a general relaxation time $\tau_R$ for
the various relaxation scenario. One can see from  Eqns. \Ref{f7},
\Ref{f8} that general relaxation behavior is defined by
complicated nonlinear interactions and combinations of short-range
and long-range relaxation processes on the first and second
relaxation levels. Therefore, the general behavior of relaxation
time $\tau_R$ is determined by a complicated combination of
relaxation rates $R_1^{(s)}$, $R_2^{(s)}$, $|\Lambda_1|^{1/2}$ and
$|\Lambda_2|^{1/2}$ on these two interconnected relaxation levels.
Calculating these
 five relaxation parameters : $\tau_R$,
$\tau_1^{(s)}=R_1^{-1}$, $\tau_2^{(s)}=R_2^{-1}$,
 $\tau_1^{(l)}=|\Lambda_1|^{-1/2}$ and $\tau_2^{(l)}=|\Lambda_2|^{-1/2}$ we can receive
a rather detailed and specific singularities of relaxation
processes in a complex systems.

\section{Results}
Figure 1 shows force output from young (B), old (C), and older-old
(D) participants at 20\% MVC at the constant (t1) and sine wave
(t2) targets. The amount of force variability was examined by
calculating the root mean square error (see, for details
\cite{Vall5}).
\begin{figure}[ht!]
\leavevmode \centering
\includegraphics[width=5.5in, height=6.0in, angle=0]{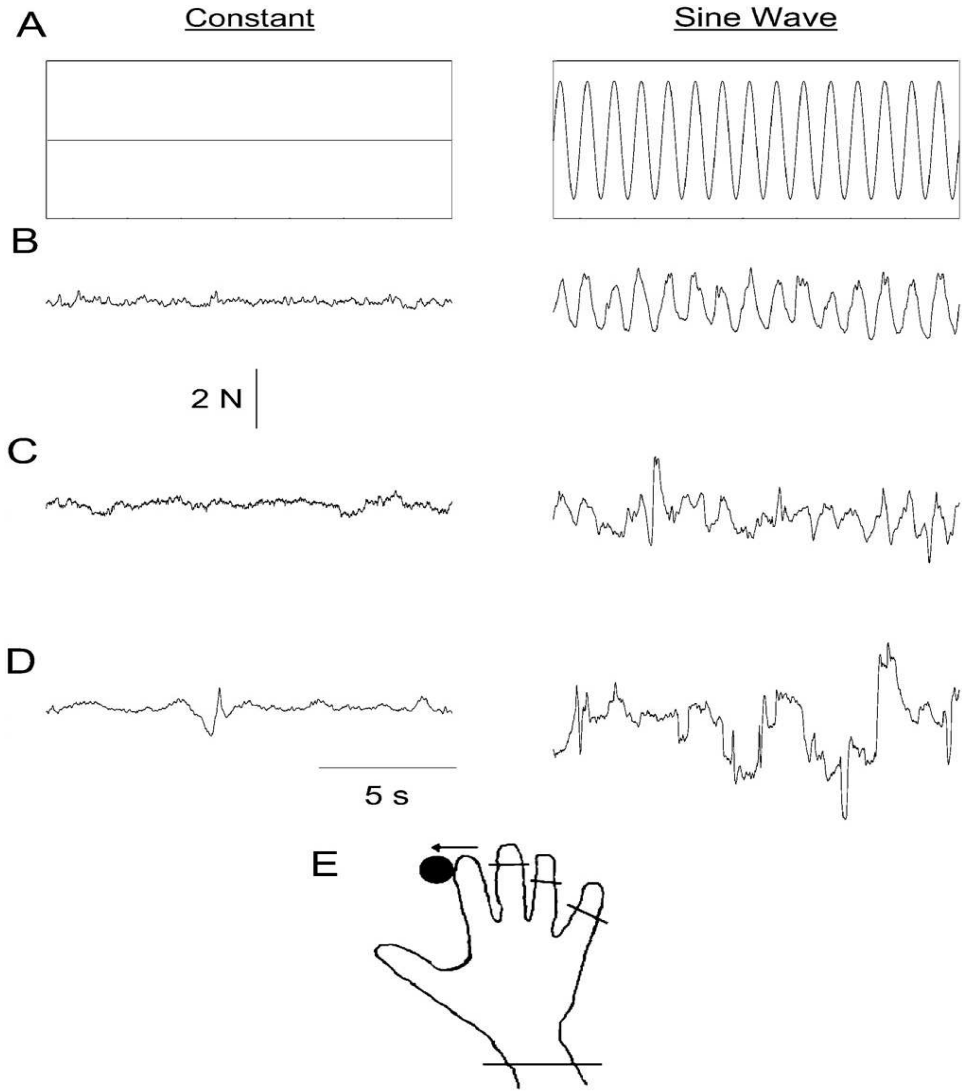}
\caption{Force output from young (B), old (C), and older-old (D)
participants at 20\% MVC at the constant (t1) and sine wave (t2)
targets.}
\end{figure}

Figure 2 depicts the values of short-range relaxation parameter on
a first relaxation level $\lambda_1$ ($R_ 1^{(s)} = |\lambda_1|$
is the corresponding relaxation rate) with following name
structure.
 \begin{center}
\begin{tabular}{p{3cm}|p{3cm}|p{3cm}}
\hline
\hline
 ~~~~1 - s1c1t1 & ~~~~~9 - s2c1t1  & ~~~~17 - s3c1t1 \\
 ~~~~2 - s1c1t2 & ~~~~10 - s2c1t2 & ~~~~18 - s3c1t1 \\
 ~~~~3 - s1c2t1 & ~~~~11 - s2c2t1 & ~~~~19 - s3c2t1 \\
 ~~~~4 - s1c2t2 & ~~~~12 - s2c2t2 & ~~~~20 - s3c2t2 \\
 ~~~~5 - s1c3t1 & ~~~~13 - s2c3t1 & ~~~~21 - s3c3t1 \\
 ~~~~6 - s1c3t2 & ~~~~14 - s2c3t2 & ~~~~22 - s3c3t2 \\
 ~~~~7 - s1c4t1 & ~~~~15 - s2c4t1 & ~~~~23 - s3c4t1 \\
 ~~~~8 - s1c4t2 & ~~~~16 - s2c4t2 & ~~~~24 - s3c4t2 \\
\hline
\hline
\end{tabular}\\
\end{center}
The file naming includes the experimental design AGE GROUP (3)
$\times$ FORCE LEVEL (4) $\times$ TRIALS (2): \\ s1, s2, s3 (group
number):  s1-young (20-24 years), s2-old (64-69 years), s3-oldest
old (75-90 years); \\ c1, c2, c3, c4  (force level): c1-5\%, c2-10
\%, c3-20 \%, c4-40 \%; \\ t1, t2 (trial number): t1-trial 1,
t2-trial 2.

Each file contains 25 $s$ force data (in Newtons) collected at 100
$Hz~(\tau=0,01 s)$. A more detailed description of the experiment
and data collection is found in \cite{Vall5}.
\begin{figure}[ht!]
     \leavevmode
\centering
\includegraphics[width=5.5in, height=6.5in, angle=90]{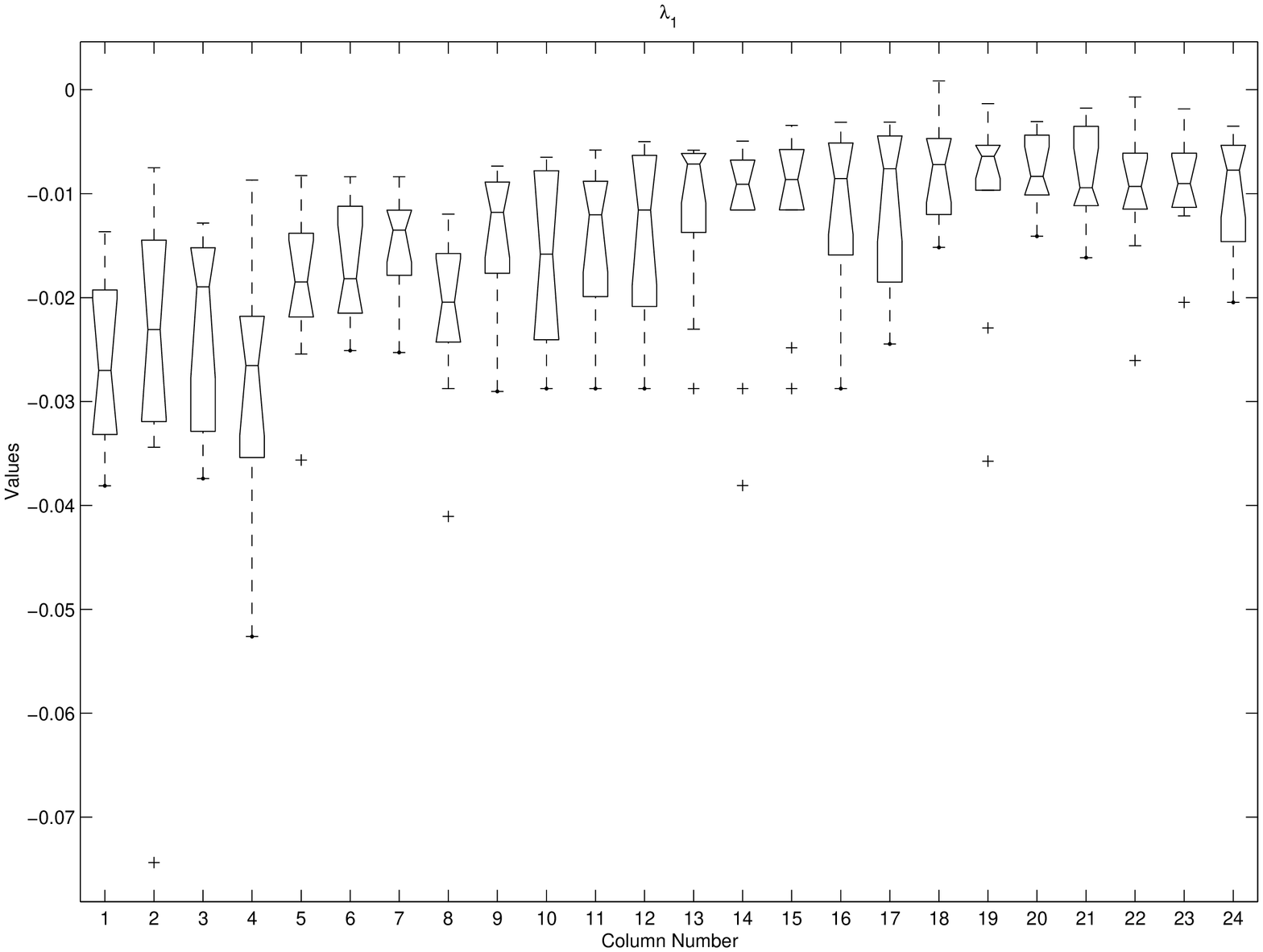}
\caption{Relaxation parameter due to short-range correlations
$\lambda_1$ (or short-range
 relaxation rate $R_1^{(s)} =|\lambda_1|$) for the all studied groups
 at first relaxation level in units of $\tau^{-1}$, where $\tau=0,01 \ s$ is
 a discretization time.
 The details of description and processing of statistical data see in text of the paper.
 Data testify for fast relaxation for young ($R_1^{(s)}=0,02 \ \tau^{-1}$),
  more slow relaxation for old ($R_1^{(s)}=0,01 \ \tau^{-1}$) and oldest old
 ($R_1^{(s)}=0,007 \ \tau^{-1}$) groups. In a whole a difference for short-range
 relaxation rate for ratio of young / old constitute 1,8 times, for young / oldest old
 is 2,66 times and for old / oldest old is 1,44 times. In a separate cases
 last ratio constitute 3,47 and more times!}
\end{figure}

Figure 2 shows numerical values of parameter $\lambda_1$ for all
three groups s1, s2 and s3. In Figure 2 the box lines at the lower
quartile, median and upper quartile values. The whiskers are lines
extending from end of the box to show the extent of the rest of
the data. Outliers are data with values beyond the ends of the
whiskers. We should add still, that first  eight boxes correspond
to s1 group, next eight boxes correspond to the second (s2) group
and last ones correspond to the third (s3) group. One can
calculate from Fig. 2 that the mean value of parameter $\lambda_1$
for first (s1) group is $\lambda_1=-0,0203~\tau^{-1}$, for second
(s2) group $\lambda_1=-0,0110~\tau^{-1}$, and for third (s3) group
of participants $\lambda_1=-0,00763~\tau^{-1}$.  Similar
distinction in rates and times of the short-range relaxation is
quite explainable with the physiological points of view. We shall
note, that the greatest distinction in  young and old groups
 achieves
 3,47 times. From physical point of view it means a more fast relaxation
of the force fluctuation for young and more slower relaxation for
old and oldest old.
\begin{figure}[ht!]
     \leavevmode
\centering
\includegraphics[width=5.5in, height=6.5in, angle=90]{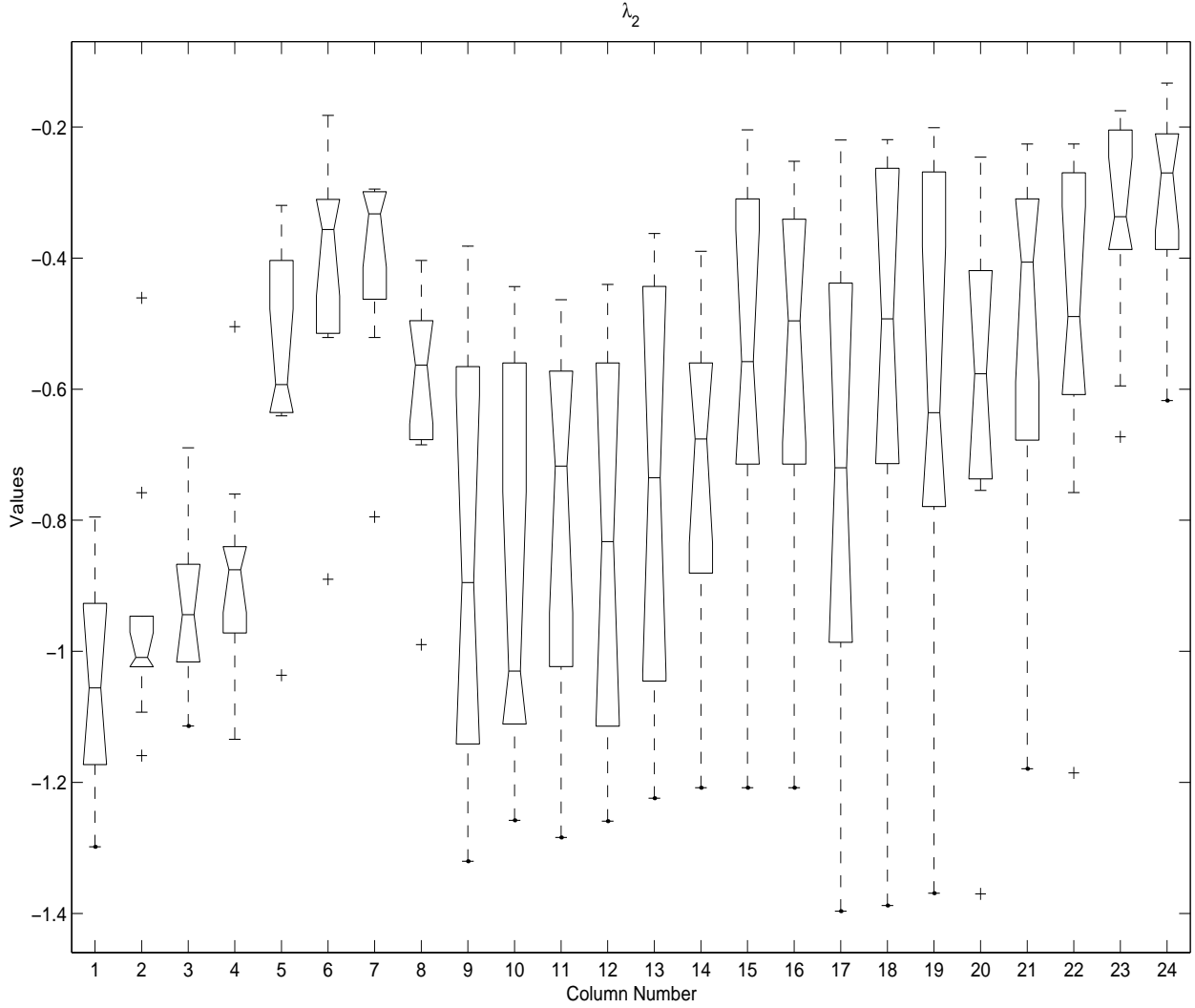}
\caption{Relaxation parameter $\lambda_2$ due to short-range
correlation
 (rate of short-range relaxation $R_2^{(s)}=|\lambda_2|$) for the all
 studied groups at second relaxation level in units if $\tau^{-1}$.
 For young group one can see a sharp decreasing of $R_2^{(s)}$ and
 $\lambda_2$ at hight  force levels (c3, c4). In the old group it is seen
a steady decreasing of relaxation rate at raising of force level
as for constant well as well for sine wave force output task. In
oldest old group steady decreasing of $R_2^{(s)}$ is observed for
constant force output task whereas for sine wave force output task
a decreasing is definite less.}
\end{figure}

Figure 3 show a similar behavior of the second short-range
relaxation parameter $\lambda_2$ for all studied groups of
participants. Authentic appreciable distinction (in 2,11 times) in
relaxation rates for  young  for force levels c1, c2 and c3, c4
pays on itself attention. Average values $R_2^{(s)} $ in units of
$\tau^{-1}$ in
 young group (0,778) and old (0,795) almost coincide with each other.
Distinctions between force levels c1, c2 and c3, c4 in group
oldest old is almost twofold. The greatest
 relaxation rate $ (R_2^{(s)}=1,056~\tau^{-1}) $ is registered in  young  group at low  force
levels (c1, c2). The least relaxation rate  is observed as for
young  group $ (R_2^{(s)}=0,50~\tau^{-1}) $ at high force levels
(c3, c4), and for oldest old group s3 $ (R_2^{(s)} =0,
36~\tau^{-1}) $ at high force levels (c3, c4).
\begin{figure}[ht!]
     \leavevmode
\centering
\includegraphics[width=5.5in, height=6.5in, angle=90]{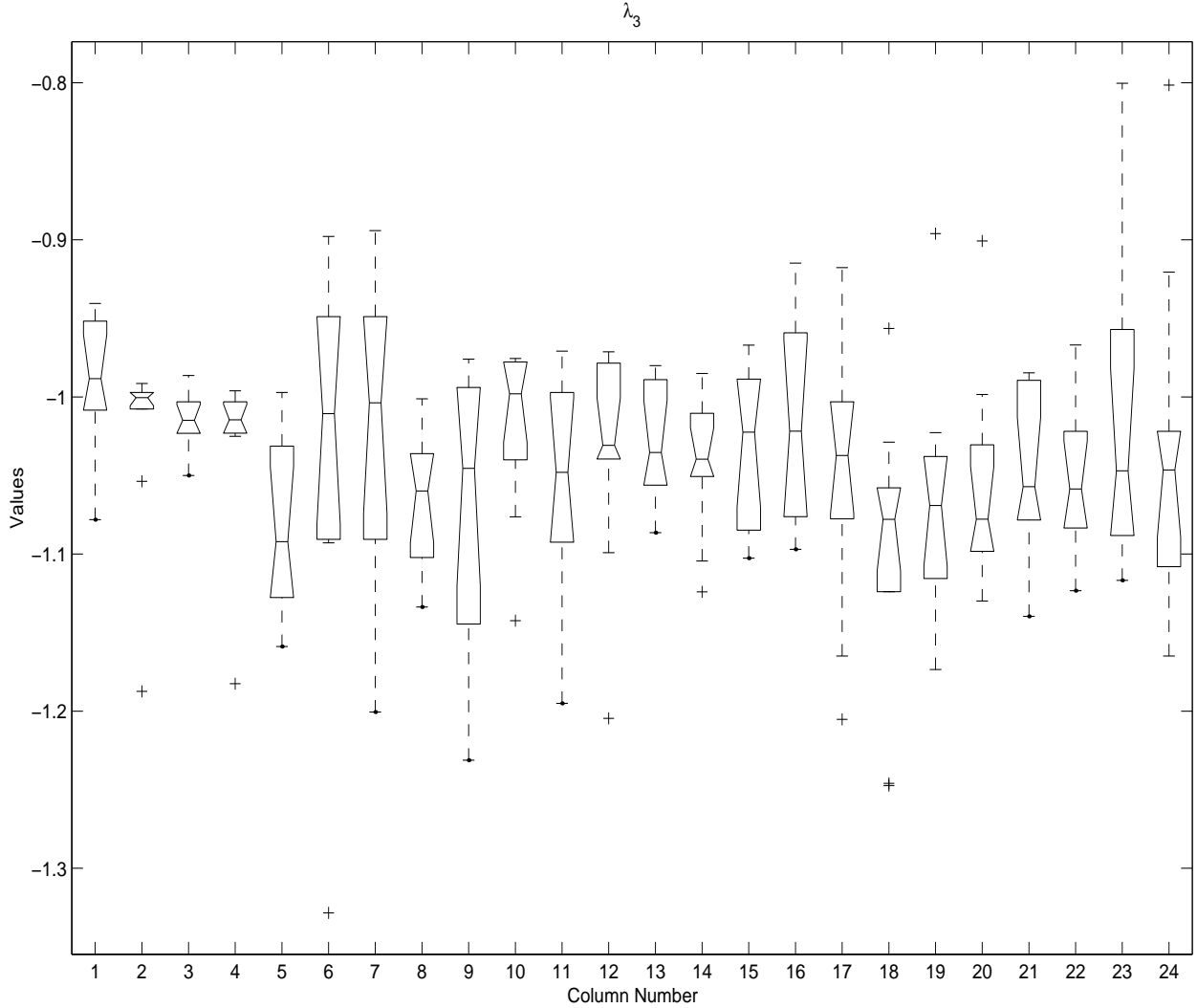}
\caption{Relaxation parameter $\lambda_3$ (rate of short-range
relaxation $R_3^{(s)}=|\lambda_3|$) due short-range correlations
for all studied groups at third relaxation level in units of
$\tau^{-1}$. One can observe almost invariable behavior of
$\lambda_3$ for a s1, s2 and s3 groups. For a young group with
enhancement of force level minimum $R_3^{(s)}$ is observed for
force level c3 (20 \%).}
\end{figure}

The behavior of the third relaxation parameter $ \lambda_3 $ is
displayed on Figure 4. Relaxation rate $R_3^{(s)} = |\lambda_3| $
on the third relaxation a level is appeared, approximately,
identical for the all three age groups.
\begin{figure}[ht!]
     \leavevmode
\centering
\includegraphics[width=5.5in, height=6.5in, angle=90]{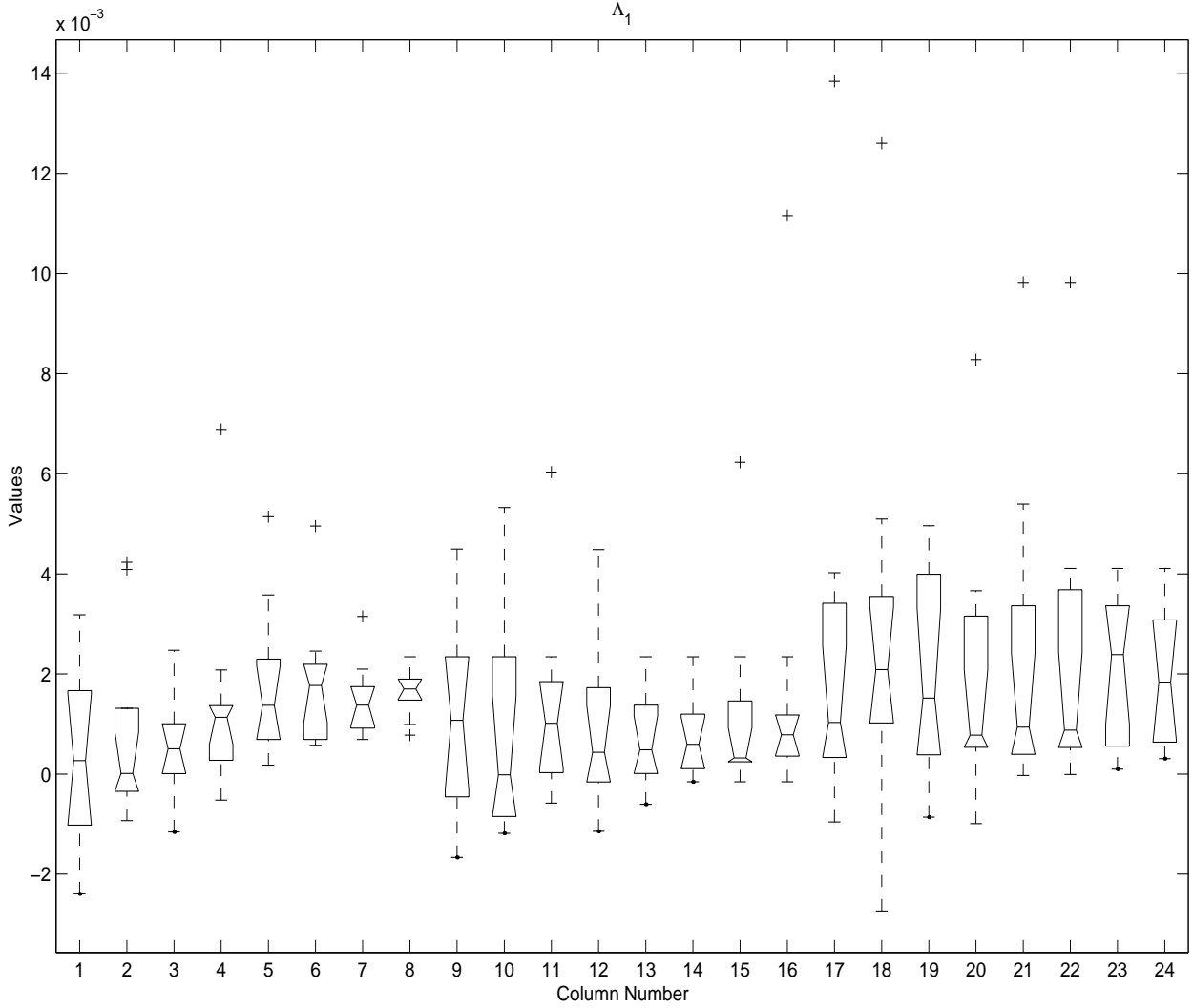}
\caption{Relaxation parameter due to long-range correlation
$\Lambda_1$ (appropriate long-range relaxation rate
$R_1^{(l)}=|\Lambda_1|^{1/2}$) for s1, s2, s3 groups at first
relaxation level in units of $\tau^{-2}$. As parameter $\Lambda_1$
has a dimension of a square of frequency, parameter
$R_i^{(l)}=|\Lambda_i|^{1/2}$ has a dimension of relaxation rate
in units of $\tau^{-1}$, where $\tau=0,01 \ s$ is discretization
time. One can see a steady increasing of this rate in four times
for young group at constant force output task and twofold
increasing for sine wave output task. For old group $R_1^{(l)}$ is
steady for various force levels, in fact. Similar behavior of
$R_1^{(l)}$ is remained for oldest old group in process of
increase of force levels.}
\end{figure}

\begin{figure}[ht!]
     \leavevmode
\centering
\includegraphics[width=5.5in, height=6.5in, angle=90]{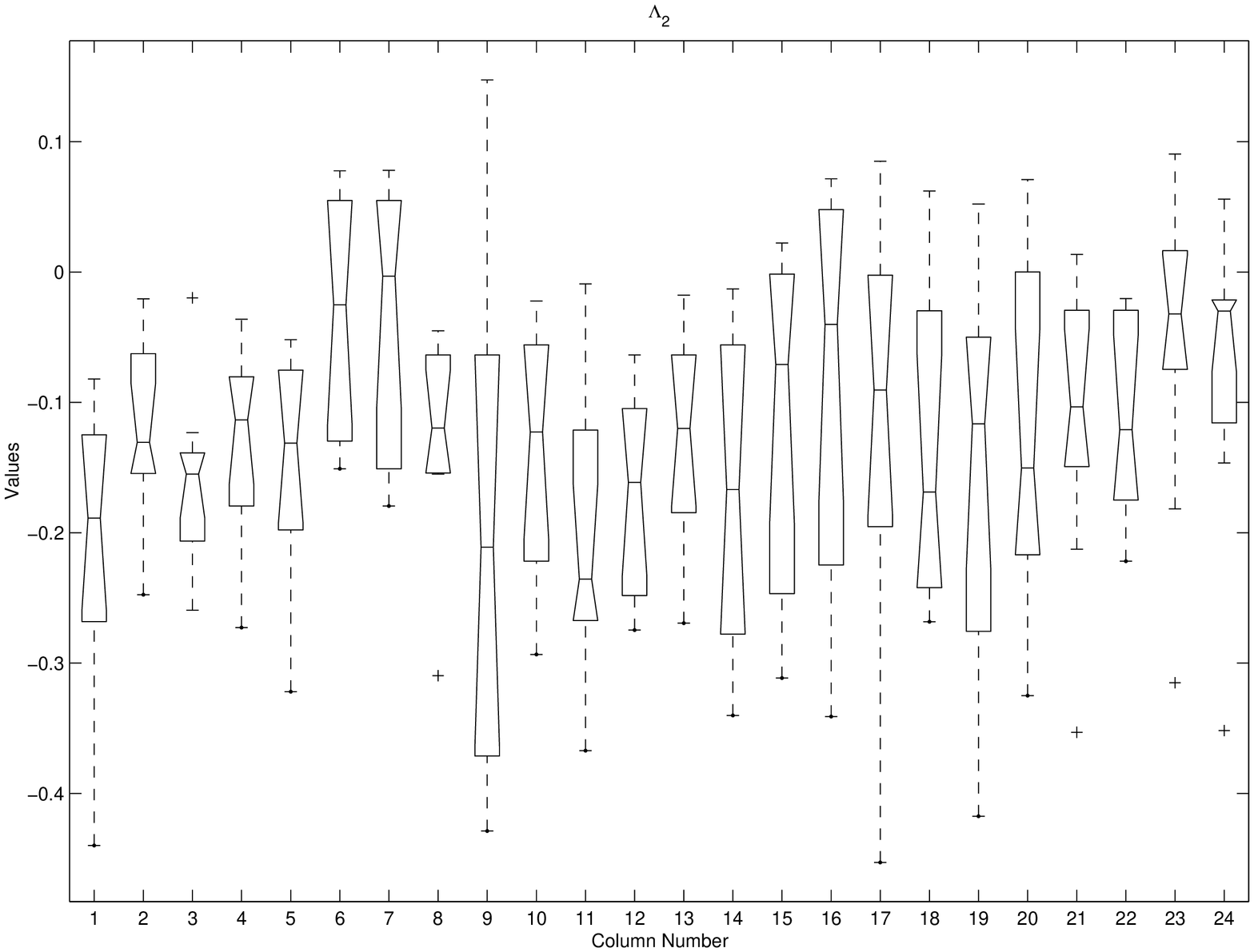}
\caption{Relaxation parameter due long-range correlation
$\Lambda_2$ (of long-range relaxation rate
$R_2^{(l)}=|\Lambda|^{1/2}$)  for all studied groups s1, s2 and s3
at second relaxation level in units of $\tau^{-2}$. One of the
specific peculiarities of this case is a negative sign of
parameter $\Lambda_2$. It means a change of relaxation mode on the
second relaxation level. For all studied groups s1, s2 and s3 we
have approximately similar relaxation behavior. Consequently from
the physical point of view one can to suppose that contribution to
relaxation due to long-range correlations is almost steady with
aging.}
\end{figure}
Relaxation parameters $\Lambda_1$ and $\Lambda_2$, connected with
long-range time correlations, are shown in Figures 5 and 6 in
units of $\tau^{-2}$. Relaxation time $ (\tau_1 ^ {(l)} = |
\Lambda_1 | ^ {1/2}) $ for the first relaxation levels are almost
identical to all three age groups (s1, s2 and s3). It testifies
that long-range correlations mechanism of force relaxation,
practically does not depend on age (average  $ \tau_1^{(l)} $ for
groups s1, s2 and s3, are equal, respectively, 0,98 $\tau$; 1,16
$\tau$ and 0,98 $\tau$). Relaxation parameter $ \Lambda_2 $ turn
out as negative for all three age groups s1, s2 and s3. This
testify to a change of relaxation mode at the transition from the
first on second relaxation levels. Relaxation times $ \tau_2 ^
{(l)} $ for groups s1, s2 and s3 changes slightly (they are equal
0,96 $\tau$ ; 0,79 $\tau$ and 1,11 $\tau$, correspondingly). It
can testify upon the weak age changes in long-range relaxation
mechanism of force output fluctuations.
\begin{figure}[ht!]
     \leavevmode
\centering
\includegraphics[width=5.5in, height=6.5in, angle=90]{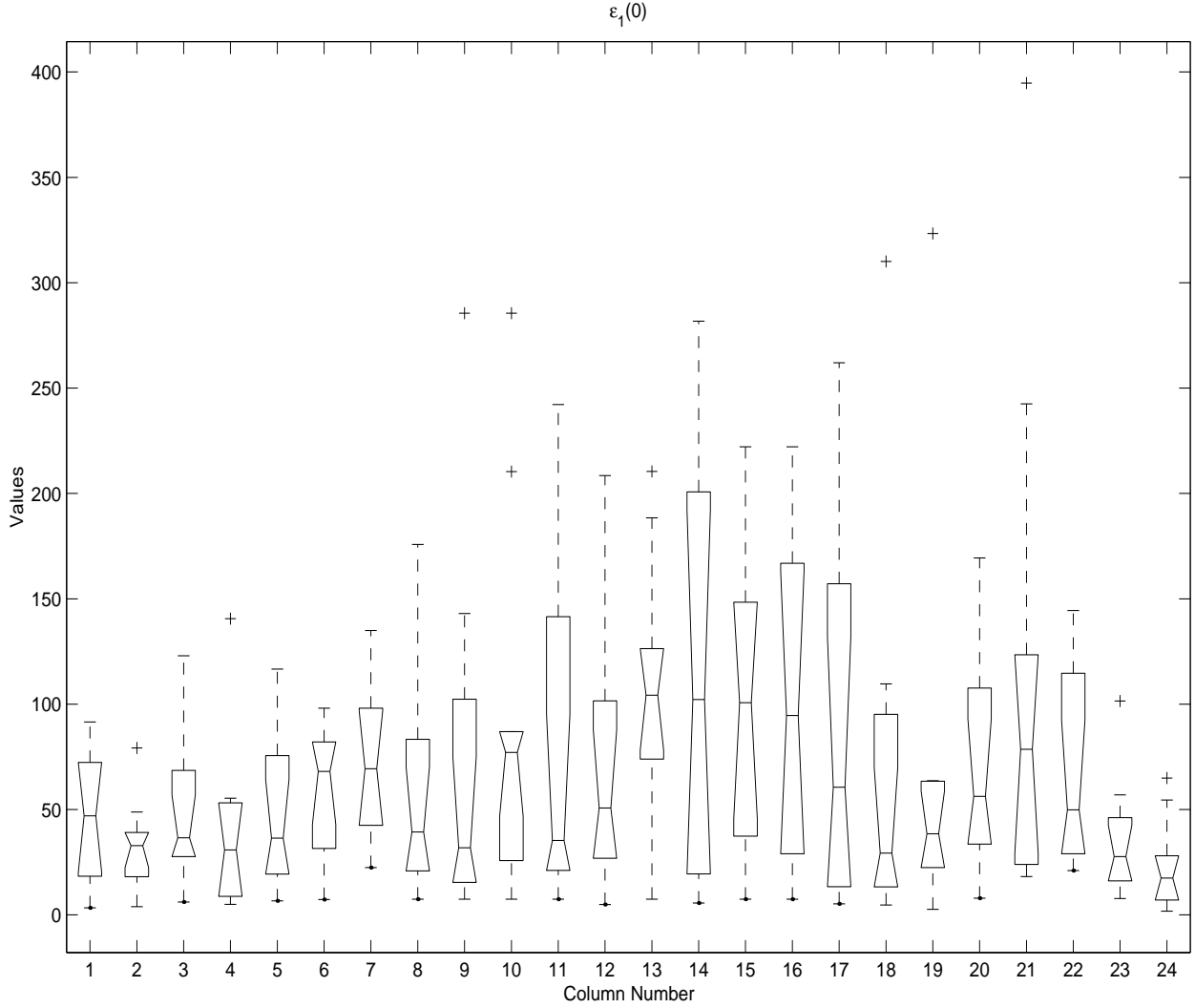}
\caption{First point of the statistical non-Markovity parameter
$\varepsilon_1(0)$ on zero frequency $\omega=0$ as the information
measure of memory for groups s1, s2 and s3. Values of
$\varepsilon_1(0)$ is approximately same for young ($\varepsilon_1
\sim 51$) and oldest old groups and $\varepsilon_1 \sim 107$ for
old group. It means a similarity of quantitative measure of memory
and randomness (existence of Markov effects) for these two groups
(young and oldest old) and increase of this measure for group s2.}
\end{figure}

In Fig. 7 one can see a values of non-Markovity parameter $
\varepsilon_1 (0) $ of the first relaxation level on zero
frequency for three age groups s1, s2 and s3. Appreciable interest
represent, that in young  group (average value $
\varepsilon_1=51$) and oldest old group ($ \varepsilon_1=55,5 $)
the chaoticity and randomness is  approximately the same.
Simultaneously it is obvious, that the greatest chaoticity
$(\varepsilon_1 =107) $ appears in group old participants s2. Fig.
8 presents statistical non-Markovity parameter on the second level
$ \varepsilon_2 (0) $ for the all age groups. One can note, that
everywhere $ \varepsilon_2 (0) $ is almost identical and equal
unity. From results of Fig. 8 it follows, that a condition of
applicability of time-scale invariance idea and approximate
equality of memory life-time on the  first and second relaxation
levels (see, Eqn. (2))  is carried out with the high degree of
accuracy. Because of the experimental data submitted on Fig. 8,
one can  use Eqns. (6) - (7) for calculation of general relaxation
time $ \tau_R $.
\begin{figure}[ht!]
     \leavevmode
\centering
\includegraphics[width=5.5in, height=6.5in, angle=90]{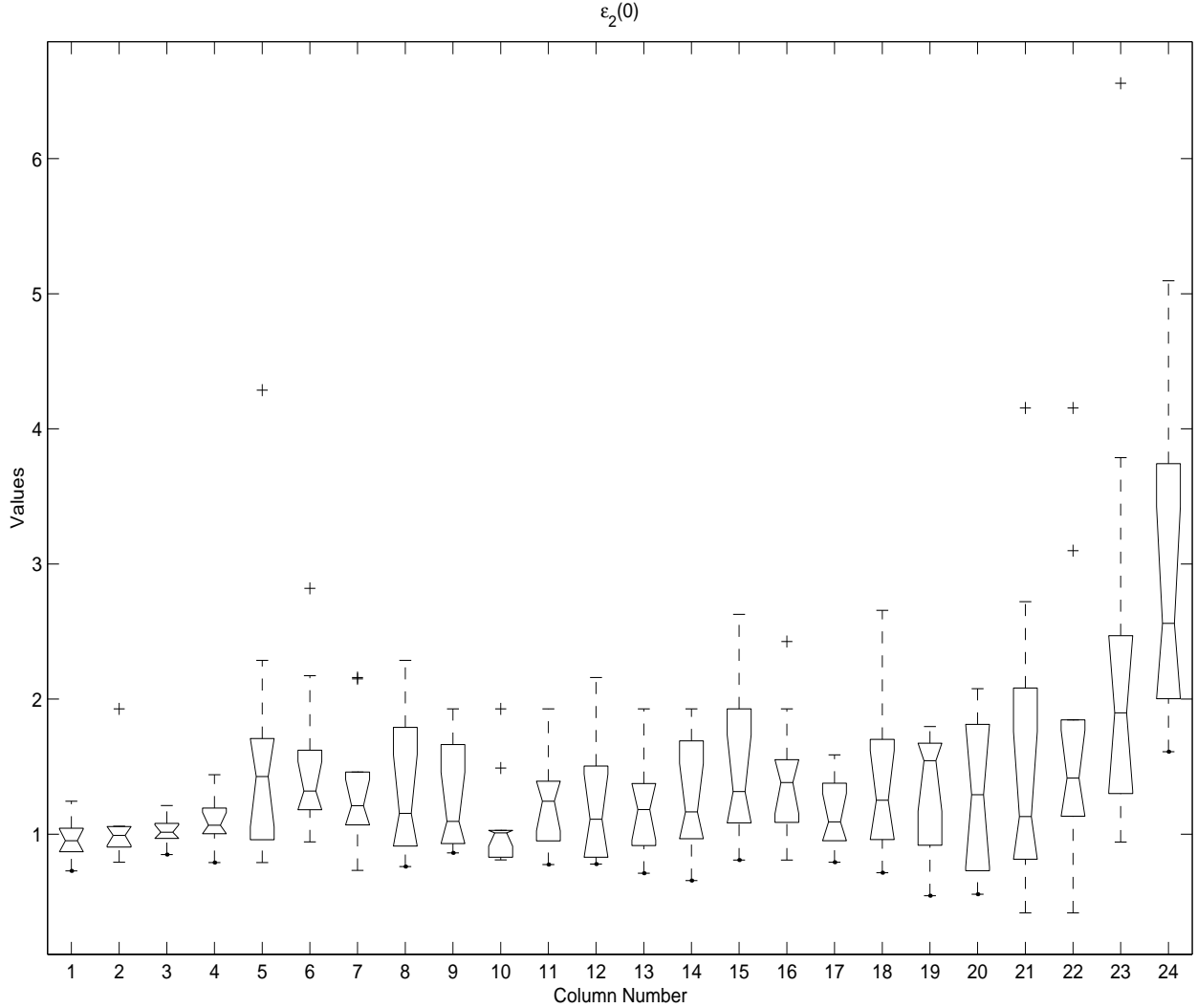}
\caption{Behavior of the second point of non-Markovity parameter
$\varepsilon_2(0)$ in all studied groups s1, s2 and s3. One can
see identical values of $\varepsilon_2(0) \sim 1$ with fine
precision. Namely similar behavior let us to use a time
scale-invariance idea and apply a solution (6)-(7) of kinetic
equations (1)-(3) for analysis of experimental data. This figure
constitute an experimental basis of our solution of kinetic
equations for TCF and MF.}
\end{figure}

\begin{figure}[ht!]
     \leavevmode
\centering
\includegraphics[width=5.5in, height=6.5in, angle=270]{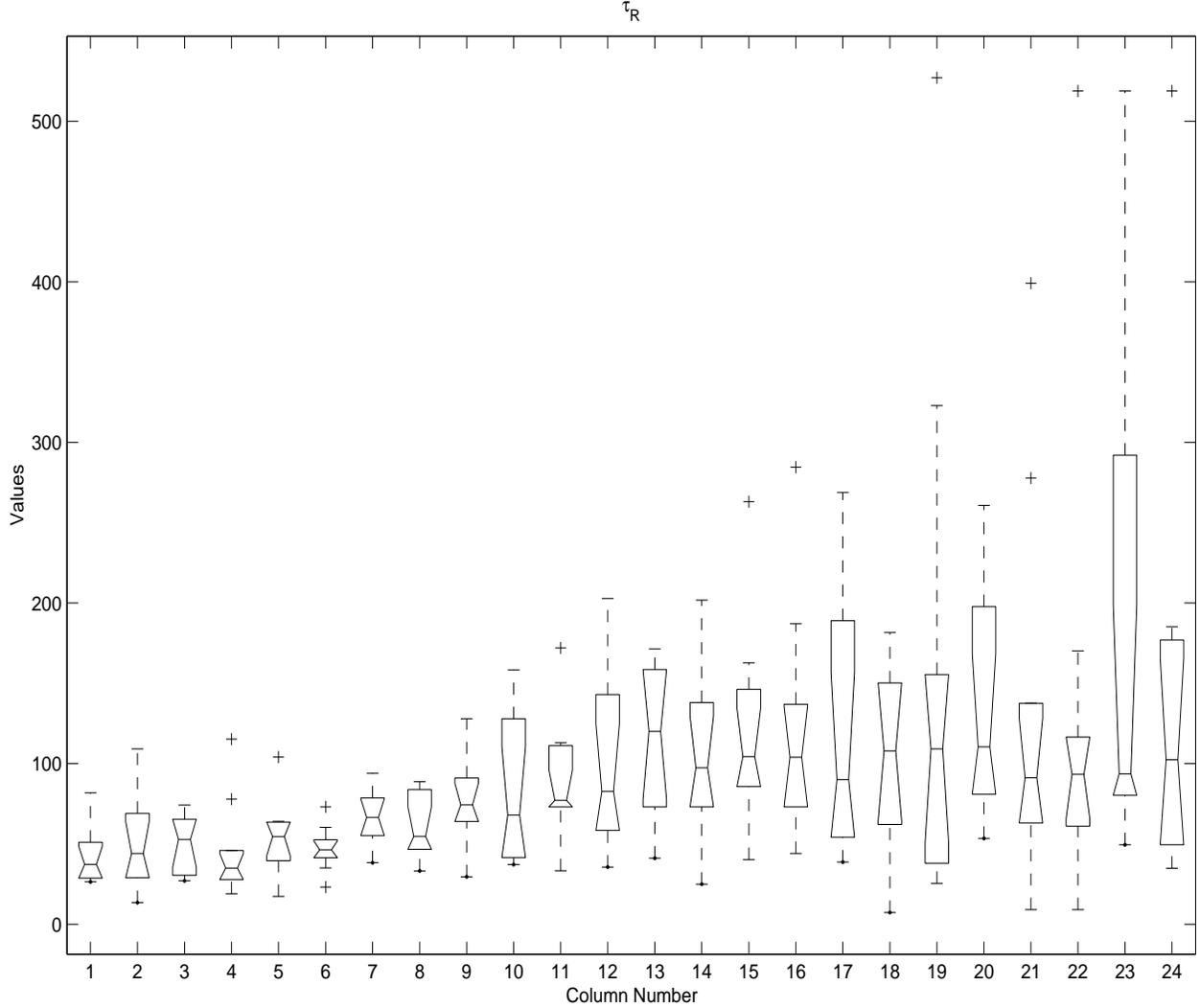}
\caption{General relaxation time $\tau_R$, calculated in
accordance with Eqns. (6), (7) of our theory for all studied
groups in units of $\tau$. Aging appears as an approximate
doubling of $\tau_R$ and slowdowningof a whole relaxation process
including as a short-range and well as a long-range correlations.}
\end{figure}

Fig. 9 presents a values of time $ \tau_R $ calculated according
our theory, Eqns. (6), (7), for all studied groups of
participants. Mean relaxation time for young group (s1) at
enhancement of force level (from c1, c2 to c3, c4) increase from
value 42,5 $ \tau $ up to value 56,07 $ \tau $. For the group of
young mean $ \tau_R $ is 49,3 $ \tau $, where as for the group of
old and oldest old mean $ \tau_R $ is equal 91,1 $ \tau $ and 99,8
$ \tau $, correspondingly. It means, that a general relaxation
time $ \tau_R $ at ageing increases on two times, approximately.
\begin{figure}[ht!]
     \leavevmode
\centering
\includegraphics[width=5.0in, height=6.5in, angle=270]{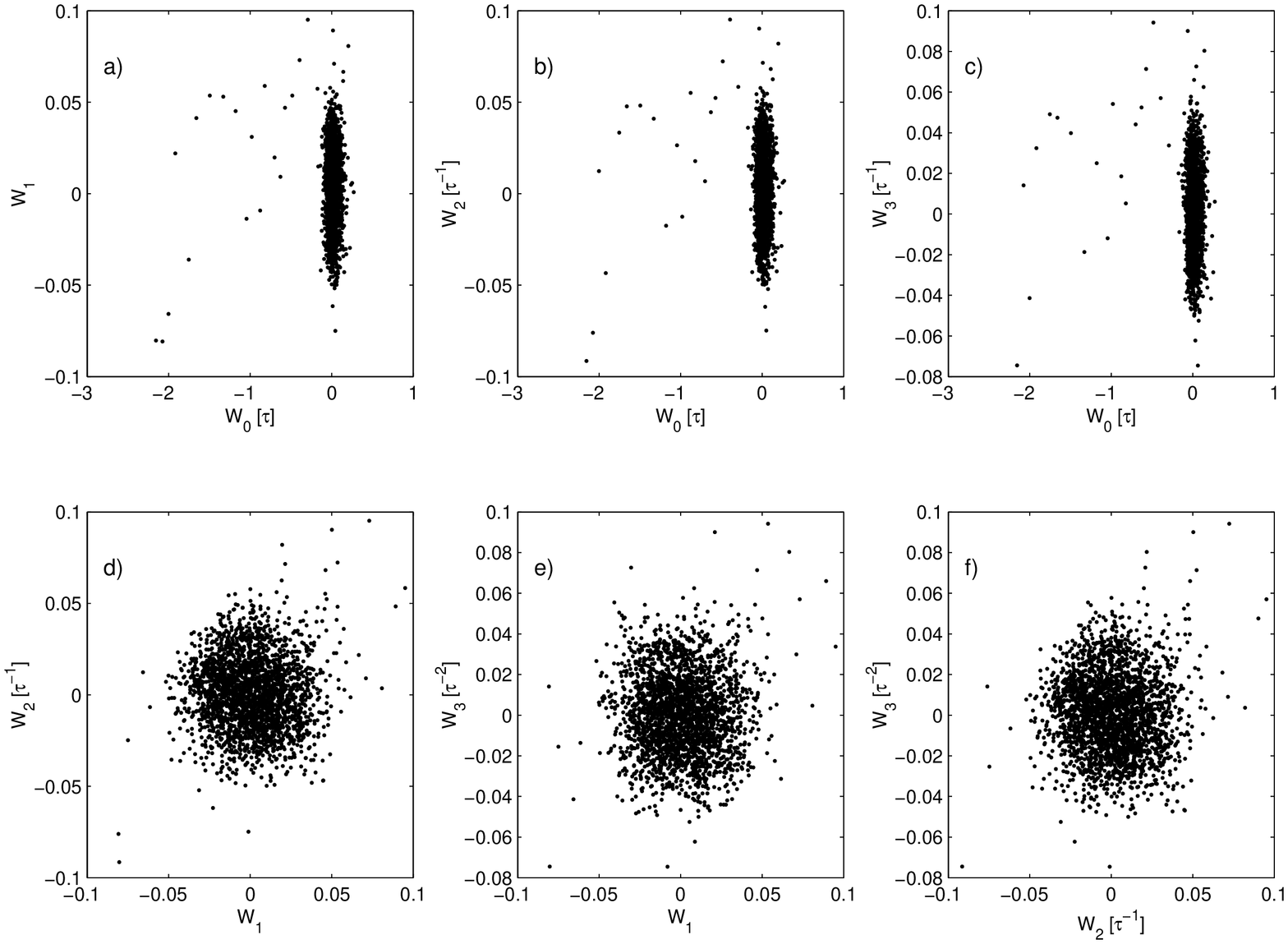}
\caption{Phase portraits in the planes ($W_i, W_j$) $i,j=1,2,3$ of
junior orthogonal dynamical variables $W_i$ for participant B4 of
young group at force level c1 and sine wave force output t2. One
can characterize these portraits by a dense nucleus and small
dissipation of phase points on the planes. Similar behavior is
typical for healthy young human.}
\end{figure}

Fig. 10 shows  the phase portraits in a planes of four junior
orthogonal dynamic variables $W_0$, $W_1$, $W_2$ and $W_3$,
calculated according theory \cite{main,eqpap,ycard} for
experimental file named s1c1t2 (B4) as an example. These portraits
appears symmetrical in planes: $(W_1,W_2)$, $(W_1,W_3)$ and
$(W_2,W_3)$. Slight deviation from this symmetry one can see for
planes of junior variables $(W_0,W_i)$, i=1,2,3. Phase clouds are
dense and concentrated.
\begin{figure}[ht!]
     \leavevmode
\centering
\includegraphics[width=5.0in, height=6.5in, angle=90]{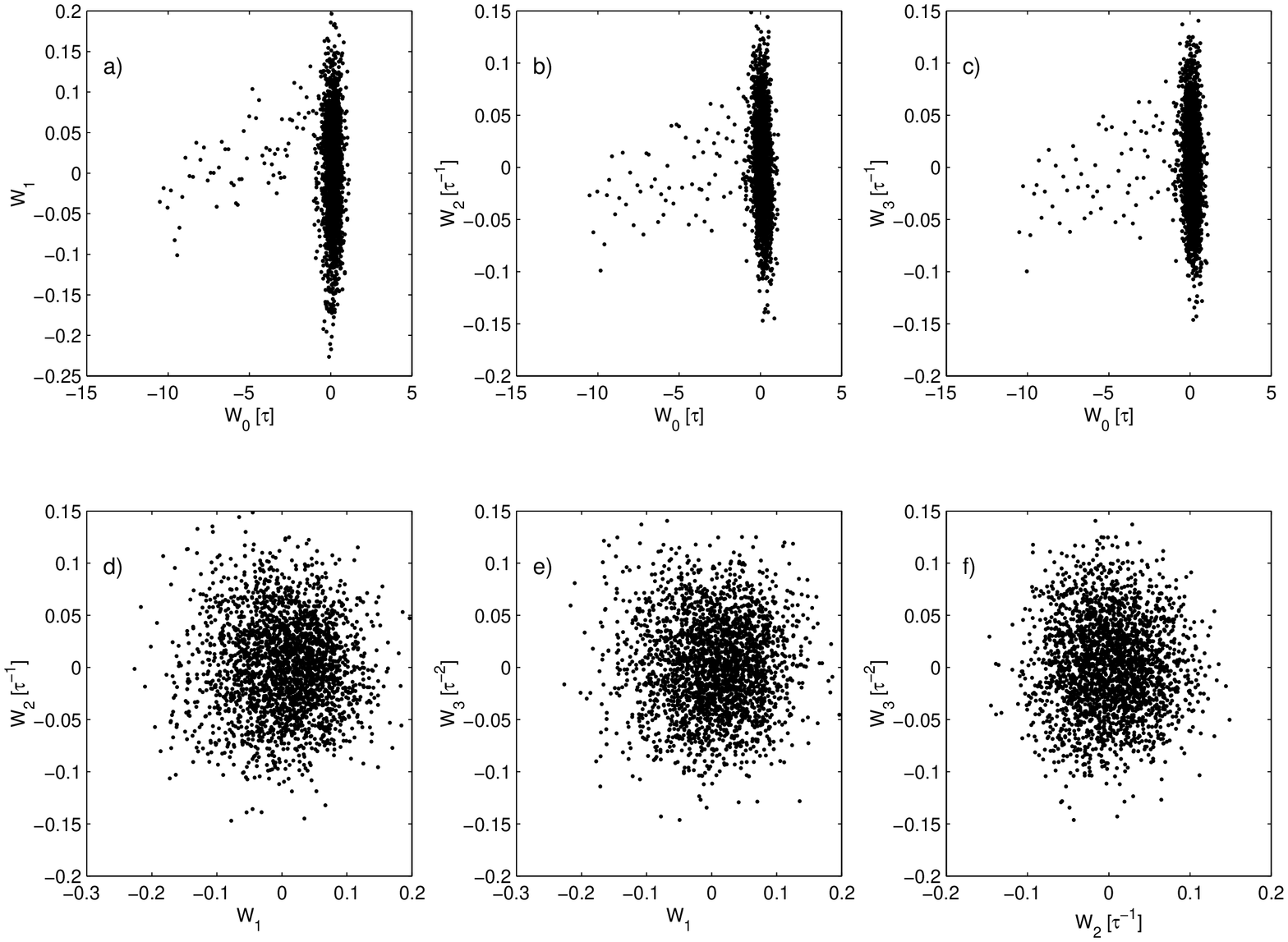}
\caption{Enhancement of force level (c4) for participant B4 from
young group reduce the scattering and appearance asymmetry of
distribution of phase points. Therefore, raising of force level
and force capacity lead to the swelling of phase clouds and to
increase of the disordering effects in phase space.}
\end{figure}
Fig. 11 present phase portraits for file name s1c4t1 of young
participant (B4) at highest force level (40 \%). One can  note  a
significant multifold swelling of volume of all phase clouds. This
testify the noticeable increasing of chaoticity of motor force
activity for this case.
\begin{figure}[ht!]
     \leavevmode
\centering
\includegraphics[width=5.0in, height=6.5in, angle=270]{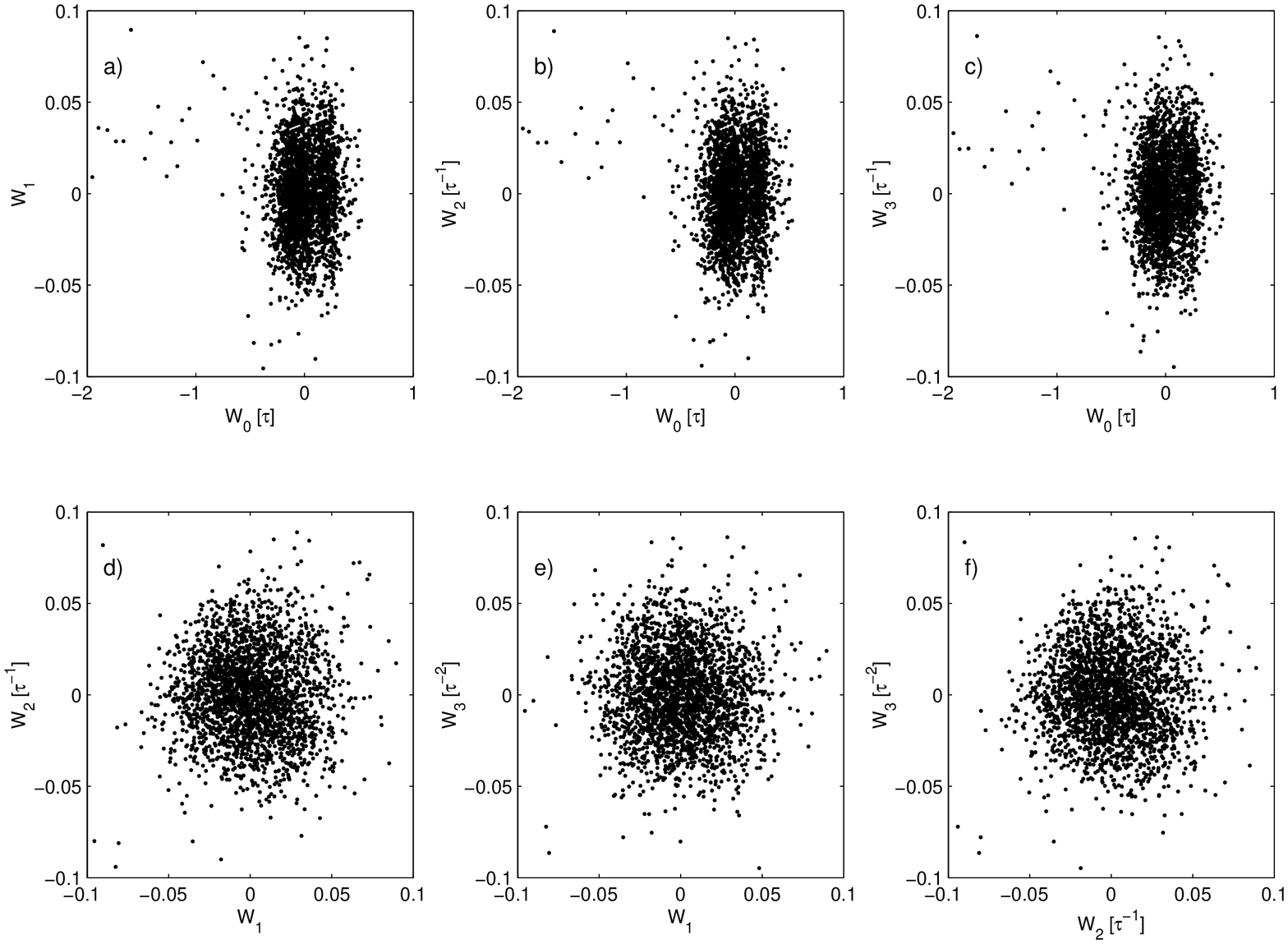}
\caption{ Enhancement of force level (c4 instead of c1) for
participant C2 from old group reduce the scattering and appearance
asymmetry of distribution of phase points. Therefore, raising of
force level and force capacity lead to the swelling of phase
clouds and to increase of disordering dynamics in phase space.}
\end{figure}

\begin{figure}[ht!]
     \leavevmode
\centering
\includegraphics[width=5.0in, height=6.5in, angle=90]{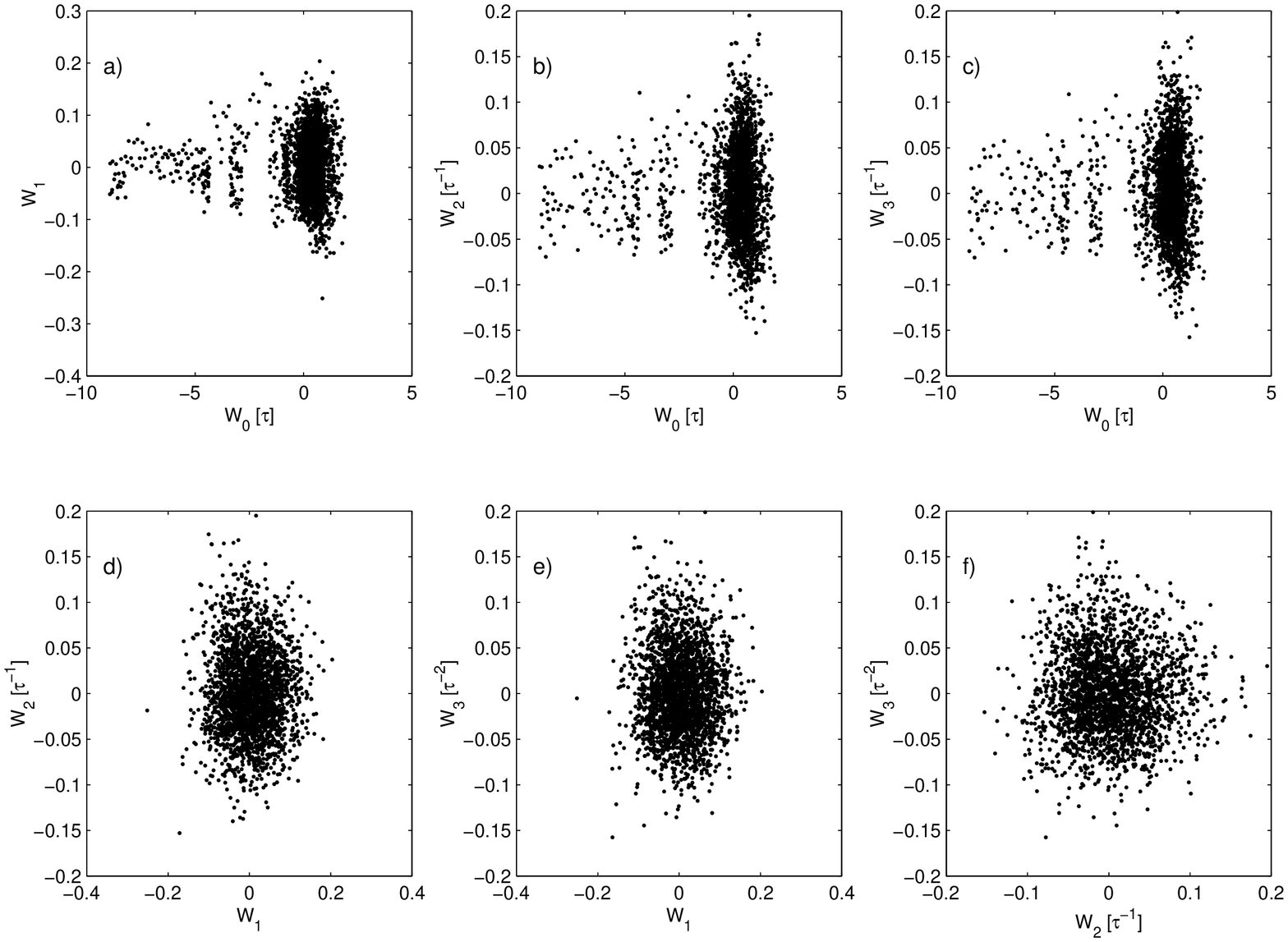}
\caption{Enhancement of force level (c4 instead of c1) for
participant C2 of old group reduce to twofold swelling of phase
clouds, remarkable stratification and scattering of phase points
in phase space.}
\end{figure}

Figs. 12 and 13 depict a change of the structure of phase clouds
for old participant (C2) for force levels c1t2 and c4t4, as an
example, correspondingly. From Fig. 12 one can find the effect of
condensation of all phase clouds. It corresponds to lowest value
of the first non-Markovity parameter $(\varepsilon_1(0) \sim 18)$
and weakly marked chaoticity of force fluctuation. At the
transition to other force level (c4, 40 \%) from Fig. 13  one can
see the remarkable increasing of the volumes of all phase clouds
and their appreciable asymmetry. It lead to increasing of
chaoticity of force fluctuation almost  7 times
($\varepsilon_1(0)=120$).

\begin{figure}[ht!]
     \leavevmode
\centering
\includegraphics[width=5.0in, height=6.5in, angle=90]{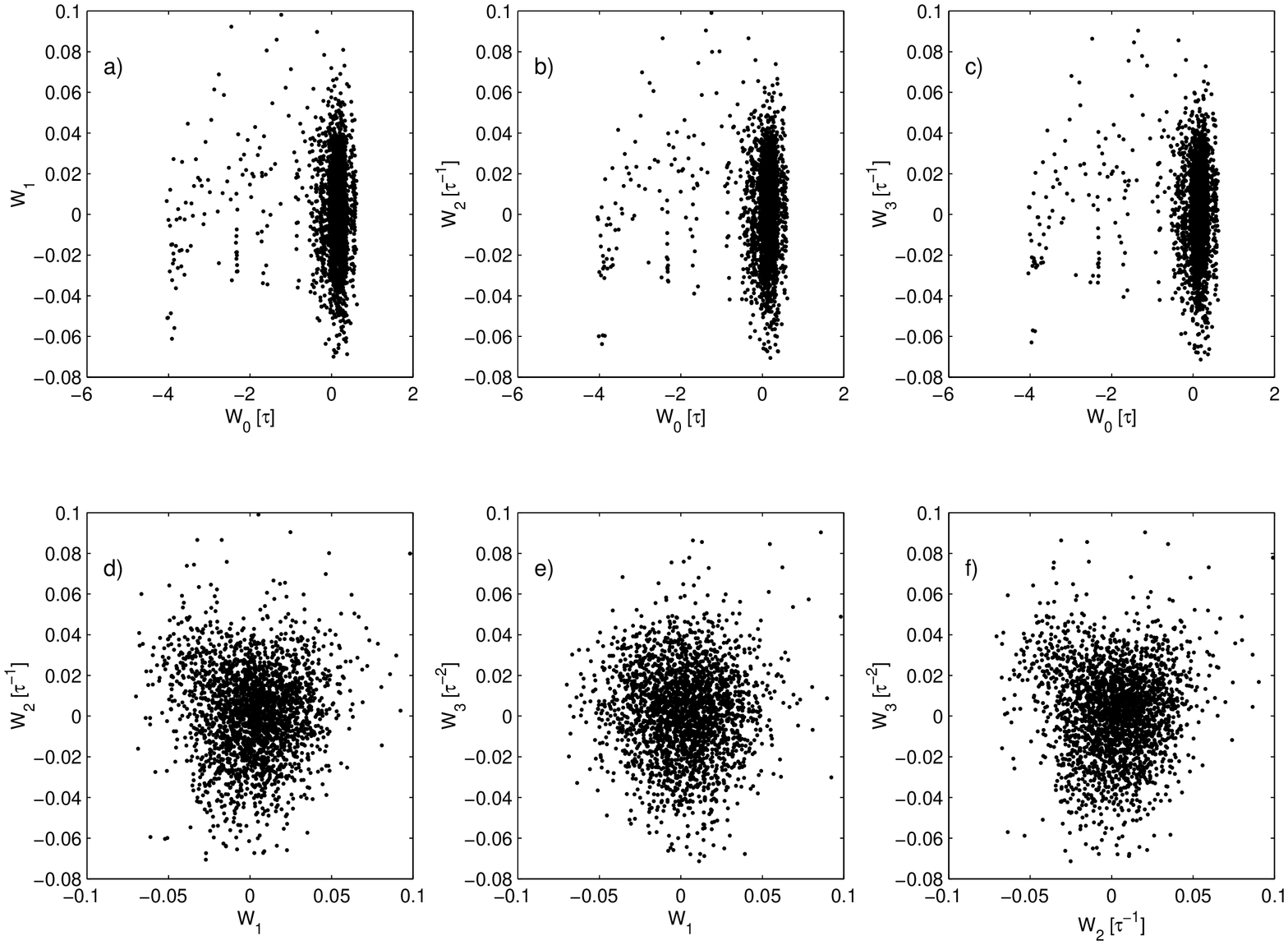}
\caption{Phase portraits in planes ($W_i, W_j$) $i,j=1,2,3$ of
junior dynamical orthogonal variables $W_i$ for participant D8 of
oldest old group at hight level force c3 (20 \%) and constant
force output task (t1). Dense phase clouds (d,e,f) are accompanied
by the slight stratification (see, figs a,b,c) of phase clouds.}
\end{figure}

Phase clouds for oldest old participants (D8) for force levels
c3t1 and c4t2 are shown on Fig. 14 and Fig. 15, correspondingly.
One can  observe a change of distribution of phase points  from
consolidated ones (for planes $(W_1,W_2)$, $(W_1,W_3)$ and
$(W_2,W_3)$)  to a more scattered distribution at transition from
force state c3t1 to state c4t2. Simultaneously, a slight
amplification of asymmetry of phase clouds in planes of junior
variables $(W_0, W_j)$, j=1,2,3 occurs. Thus, a quantitative
measure of memory $\varepsilon_1(0)$ on the first relaxation level
does not varies almost (it equal 24 and 27, correspondingly).
Therefore, the degree of Markovization and chaoticity for these
two cases does  not change.
\begin{figure}[ht!]
     \leavevmode
\centering
\includegraphics[width=5.0in, height=6.5in, angle=90]{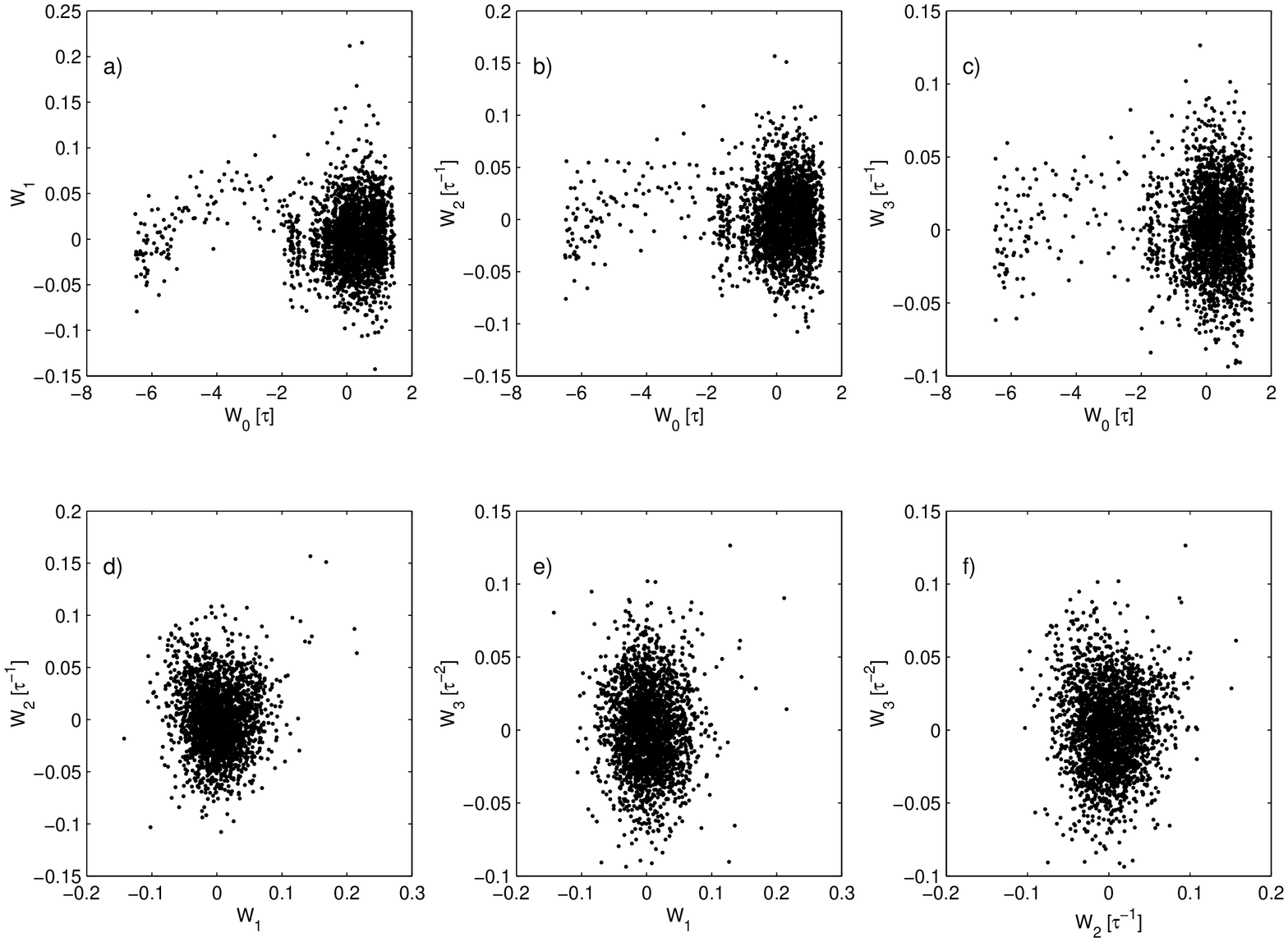}
\caption{ Twofold enhancement of force level (c4, 40 \% instead of
c3, 20 \%) for participant D8 of oldest old group at sine wave
force task reduce on minor change of the structure of phase
portraits and phase clouds without any stratification and
scattering of phase points in a phase space.}
\end{figure}

\begin{figure}[ht!]
     \leavevmode
\centering
\includegraphics[width=5.0in, height=6.5in, angle=270]{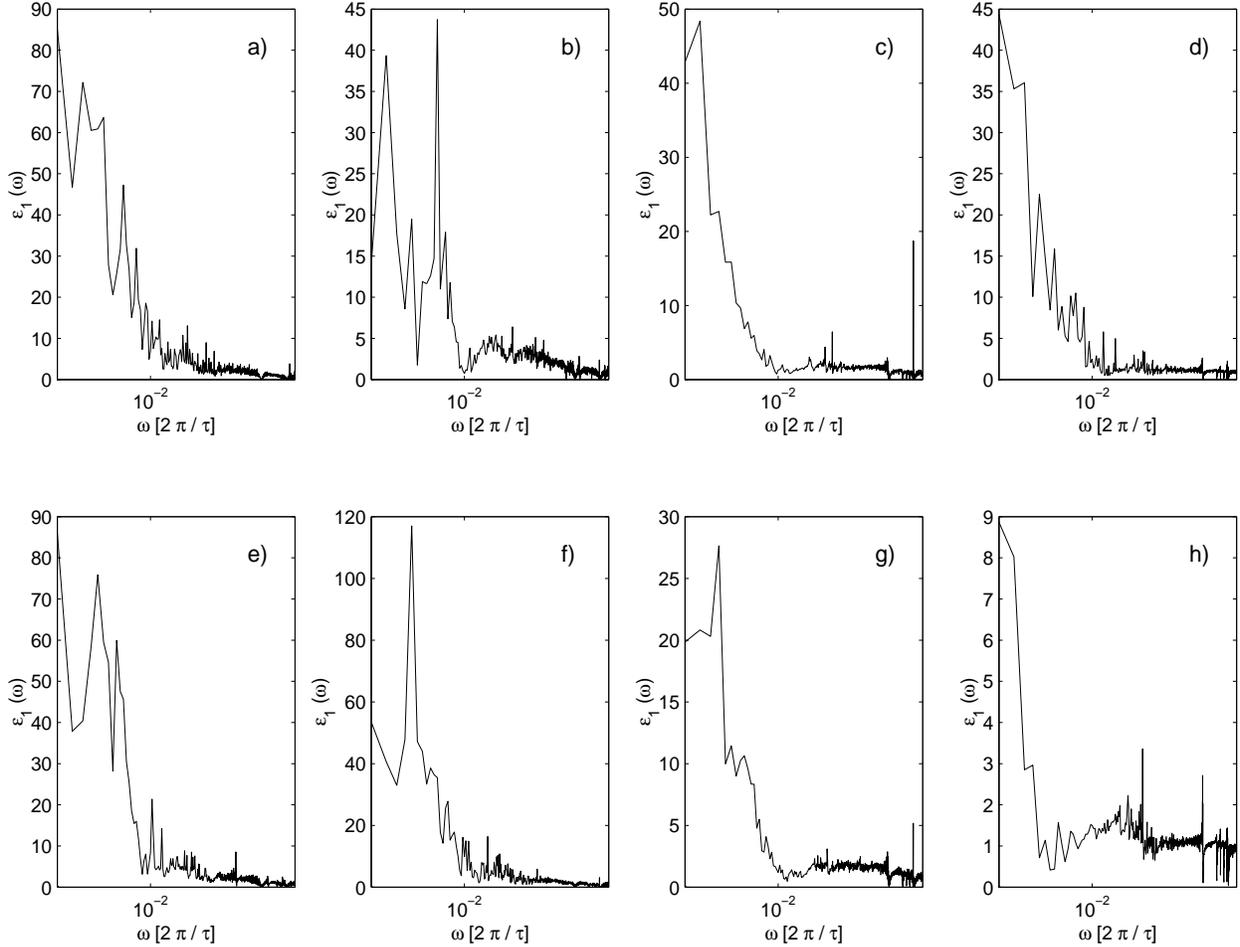}
\caption{Frequency dependence of the first point of non-Markovity
parameter $\varepsilon_1(\omega)$ for participant B5 of young
group for 8 trials: at four force levels (c1, c2, c3, c4) and at
two form (t1, t2) of force output task. Data demonstrate almost
invariable behavior of $\varepsilon_1(\omega)$ at constant force
output task, and steady decreasing of quantitative measure of
memory at sine wave force output task (for example,
$\varepsilon_1(0)$ decrease from value of 40 to value of 8). }
\end{figure}

\begin{figure}[ht!]
     \leavevmode
\centering
\includegraphics[width=5.0in, height=6.5in, angle=270]{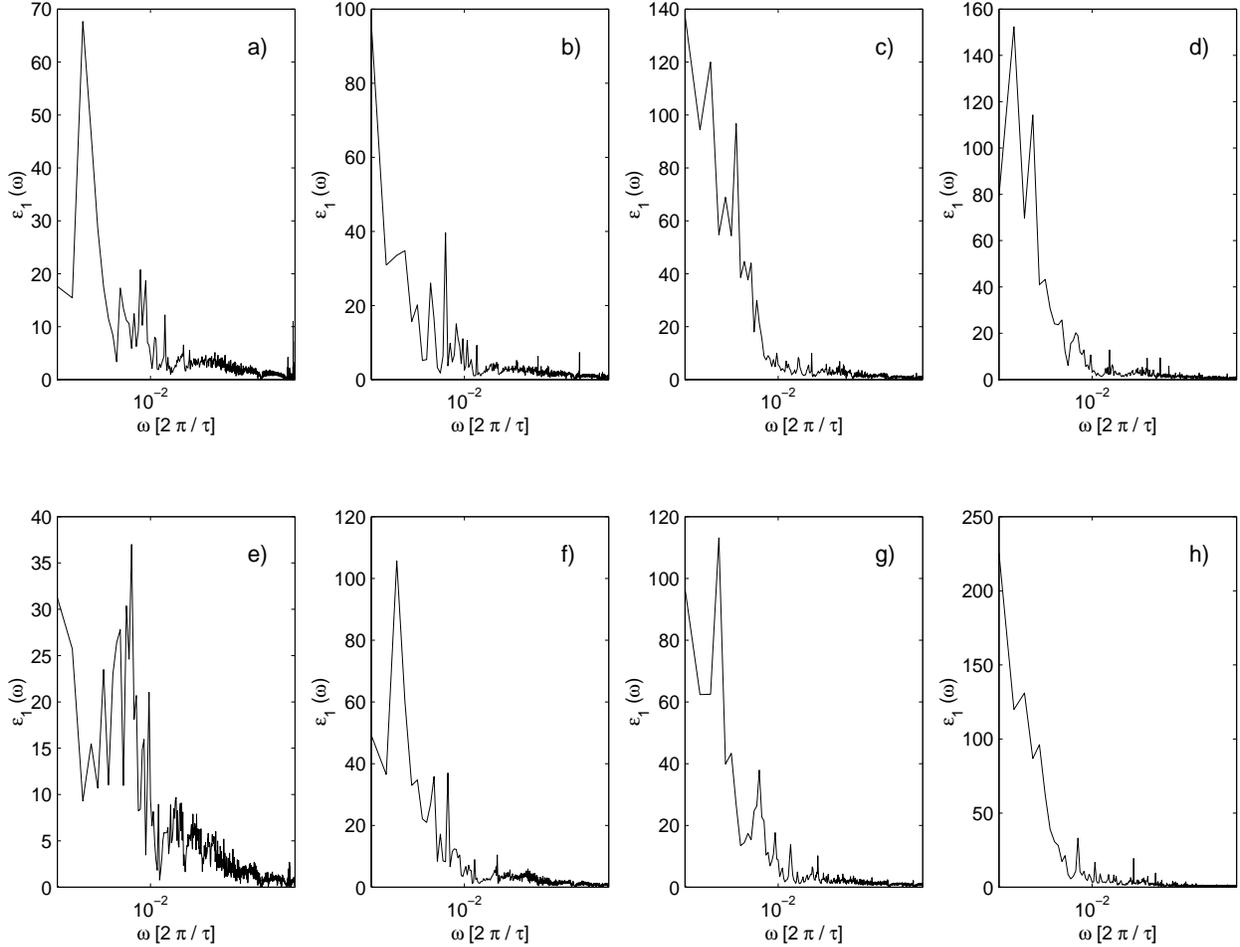}
\caption{Frequency dependence of the first point of non-Markovity
parameter $\varepsilon_1(\omega)$ for participant C2 of old group
for 8 trials: for force levels (5 \%, 10 \%, 20 \% and 40 \%) and
two form (t1, t2) of force output task. Values of quantitative
measure of memory ($\varepsilon_1(0)$) increase almost at ten
times at constant force output task, whereas it increase almost at
five times at sine wave force output task.}
\end{figure}

\begin{figure}[ht!]
     \leavevmode
\centering
\includegraphics[width=5.0in, height=6.5in, angle=270]{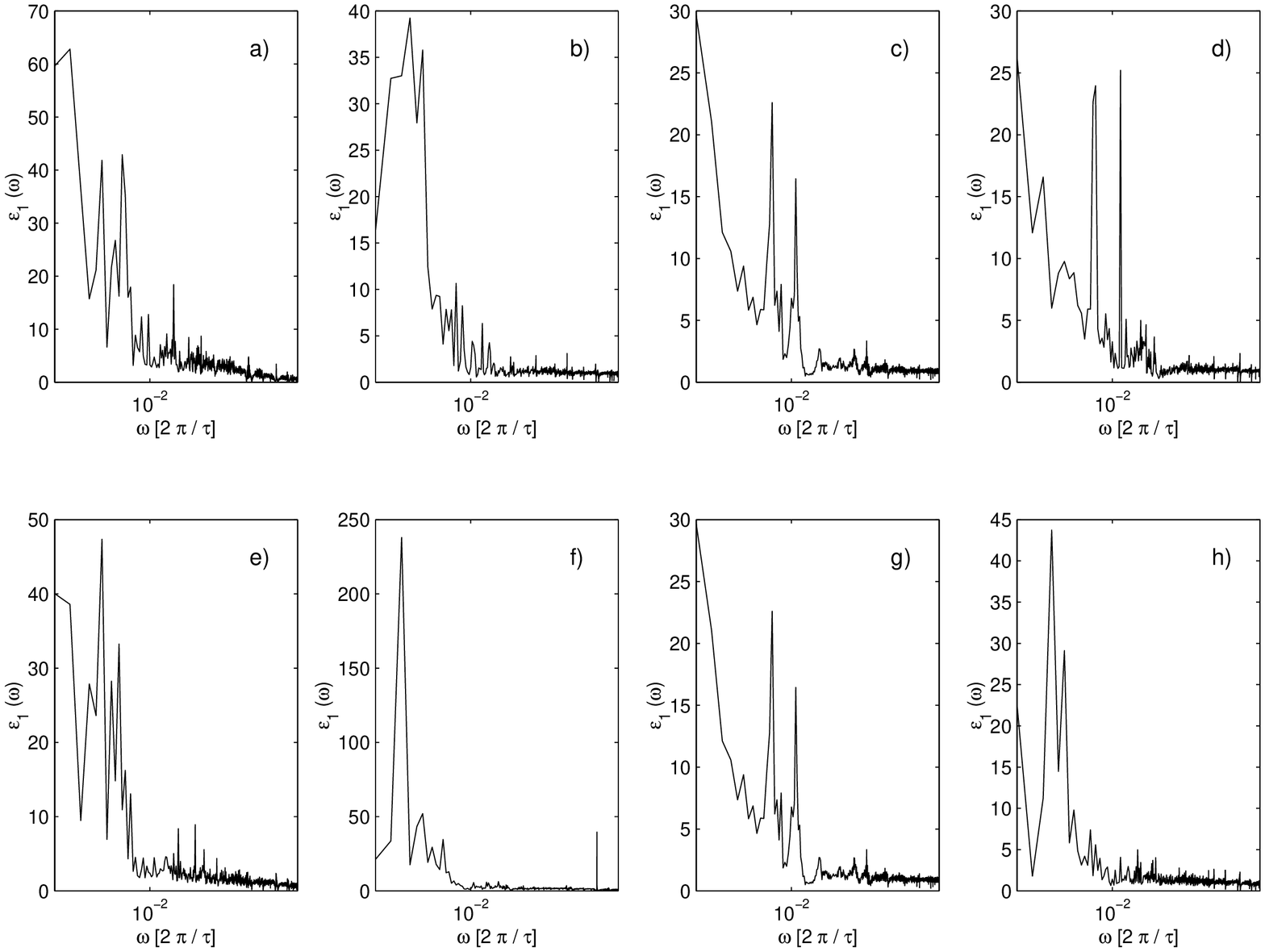}
\caption{ Frequency dependence (a,b,c,d) of the first point of
non-Markovity parameter $\varepsilon_1(\omega)$ for participant D2
of the oldest old group is related with the steady decreasing of
quantitative measure of memory and randomness at enhancement of
force level (5 \%, 10 \%, 20 \% and 40 \%) and constant force
output task (t1). Similar behavior is observed  for sine wave
force output task (t2) also. Simultaneously, one can see low
frequency range (from zero up to $10^{-2} Hz)$ with the big values
of $\varepsilon_1(\omega)$, that testify the existence a steady
frequency region of Markov random effects.}
\end{figure}

Figs. 16,17 and 18 demonstrate a frequency dependence of the first
non-Markovity parameter $\varepsilon_1(\omega)$ for young (B5) old
(C2) and oldest old (D2) participants in all force levels as an
examples. From Fig. 16 for participant D5 one can see that Markov
effect at ultralow frequency remain constant at force level t1
(constant). At sine wave (e,f,g,h) force output task one can
observe a progressive and steady decline of $\varepsilon_1(0)$
(from value of 50 on value of 8). It is connected with a reduction
of Markov effects and steady amplification of non-Markov effects.
Similar behavior of $\varepsilon_1(\omega)$ reflects the
appearance of slight robustness in force fluctuation for young
participants in process of increasing of force level. A data for
old (C2) presents opposite example, as can one see from Fig. 17.
We see a drastic amplification of Markov effects in process of
enhancement of force level for constant (a,b,c,d) and sine wave
(e,f,g,h) force output task. This means a steady Markovization of
a process, that is, a steady amplification of Markov random
effects. From Fig. 18 one can note analogous behavior of
$\varepsilon_1(\omega)$ for oldest old (D2) at the constant
(a,b,c,d) and sine wave (e-h) force output task. It testifies
about a gradual and more distinctive display of non-Markov effects
and robustness in force behavior for participant (D2).

\section{Discussion and conclusion}
The purpose of this study was to examine the structure in the time
and frequency domains of force output variability as a function of
human aging.

For the analysis of force fluctuation we have used the statistical
relaxation singularities of motor system related to the
short-range and long-range time correlation.

It is necessary to emphasize some fundamental points, following
from the study. Force output dynamics for all three groups of
healthy (young, old and oldest old) can be characterized rather a
high level of chaoticity and randomness on the first relaxation
level of force output fluctuation. For all files studied we have
found out, that $ \varepsilon_1 (\omega) $ changes in a wide
interval of values (7-300). It allows to state a strong
chaotization and sharp expressed Markov effects for all healthy
groups. As is known, beginning of illnesses, for example Parkinson
disease (see, for instance, \cite {PhysaGait}) results to
suppression of Markov  effects and appearance  of non-Markovity
effects and robustness.

The same level of a randomness observed for young $ (\varepsilon_1
\sim 51) $ and oldest old $ (\varepsilon_1 \sim 56) $  turned out
to be unexpected. Thus, on average the rather highest level of a
randomness $ (\varepsilon_1 \sim 107) $ are registered in old
group. These values of NMR $ (\varepsilon_1) $  signify healthy
state for studied participants. Therefore it is possible to state
with assurance that the information measure of mamory is the
greater for old, but it is  smaller for young and oldest old
groups. Steady non-Markovity $ (\varepsilon_2 \sim 51) $ on the
second relaxation level for all three age groups confirms our
supposition for the advantage of time-scale invariance idea of
relaxation processes of force output fluctuation on these level.

The relaxation singularities of force output fluctuations consist
in the following. At the first relaxation level a remarkable
distinction of relaxation rates of force output fluctuation is
observed. It means that relaxation related to the short-range
correlation is a more fast at low force level. With the
enhancement of force levels the relaxation is rather decelerated.
Noticeably more slow relaxation  appears in two  old groups s2 and
s3. Hence we can conclude that aging becomes apparent as the
notably slowing-down of relaxation processes, connected  with the
effect off the short-range correlation. In the second relaxation
level contribution of short-range correlation for all age groups
are the same. At that we notice stably (more than 2 times)
decreasing of the relaxation rate with increasing of force level
(from c1,c2 to c3,c4).

The following fact attract our steadfast attention. One can note,
that  contribution of long-range correlations in relaxation rates
is, approximately, identical for the all three age groups and  it
does not depend on age on the first relaxation level. However, in
young group here specific features are observed. They are that
here relaxations rates increase, on the average, at 1,5 times at
hight force levels (c3 and c4). On the second relaxation level the
similar picture is kept.

For analysis of experimental data we have developed  here the
statistical theory of relaxation of force output fluctuation with
taking into account: first two relaxation levels and effects of
two relaxation channels. One of the relaxation channel contains
the contribution of short-range correlation whereas other
component of relaxation reflects the effect of long-range
correlation. The analysis of experimental data shows, that one can
determine the general behavior of relaxation processes as a whole
by a complicated combination and nonlinear interaction of these
two above stated relaxation processes.

\section {Acknowledgments}

This work was partially supported by the RFBR (Grant no.
05-02-16639-a), Grant of Federal Agency of Education of Ministry
of Education and Science of Russian Federation (Grant no. RNP
2.1.1.741). This work has been supported in part (P.H.) by the
German Research Foundation, SFB-486, project A10.

\end {document}